    \documentclass[iop]{emulateapj}

\usepackage{amssymb}
\usepackage{amsmath}
\usepackage{color}

\newcommand{\etal}{et~al.}
\newcommand{\snia}{SN~Ia}
\newcommand{\sneia}{SNe~Ia}
\newcommand{\kms}{km~s$^{-1}$}
\newcommand{\wl}{$\lambda$}

\newcommand{\synapps}{\texttt{SYNAPPS}}
\newcommand{\phoenix}{\texttt{PHOENIX}}
\newcommand{\sedona}{\texttt{SEDONA}}
\newcommand{\cmfgen}{\texttt{CMFGEN}}
\newcommand{\synow}{\texttt{SYNOW}}

\shorttitle{Carbon in SNe Ia}
\shortauthors{Thomas et~al.}

\begin{document}

\title{\vspace{-0.45in}Type Ia Supernova Carbon Footprints}

\author
{
    R.~C. Thomas,\altaffilmark{1}
    G.~Aldering,\altaffilmark{2}
    P.~Antilogus,\altaffilmark{3}
    C.~Aragon,\altaffilmark{2}
    S.~Bailey,\altaffilmark{2}
    C.~Baltay,\altaffilmark{4}
    S.~Bongard,\altaffilmark{3}
    C.~Buton,\altaffilmark{5}
    A.~Canto,\altaffilmark{3}
    M.~Childress,\altaffilmark{2,6}
    N.~Chotard,\altaffilmark{7}
    Y.~Copin,\altaffilmark{7}
    H.~K. Fakhouri,\altaffilmark{2,6}
    E.~Gangler,\altaffilmark{7}
    E.~Y. Hsiao,\altaffilmark{2}
    M.~Kerschhaggl,\altaffilmark{5}
    M.~Kowalski,\altaffilmark{5}
    S.~Loken,\altaffilmark{2,*}
    P.~Nugent,\altaffilmark{1}
    K.~Paech,\altaffilmark{5}
    R.~Pain,\altaffilmark{3}
    E.~Pecontal,\altaffilmark{8}
    R.~Pereira,\altaffilmark{7}
    S.~Perlmutter,\altaffilmark{2,6}
    D.~Rabinowitz,\altaffilmark{4}
    M.~Rigault,\altaffilmark{7}
    D.~Rubin,\altaffilmark{2,6}
    K.~Runge,\altaffilmark{2}
    R.~Scalzo,\altaffilmark{9,10}
    G.~Smadja,\altaffilmark{7}
    C.~Tao,\altaffilmark{11,12}
    B.~A. Weaver,\altaffilmark{13}
    C.~Wu\altaffilmark{14}
    (The Nearby Supernova Factory);
    P.~J.~Brown,\altaffilmark{15} \& 
    P.~A.~Milne\altaffilmark{16}
    \vspace{-0.05in}
}

\altaffiltext{1}
{
    Computational Cosmology Center, Computational Research Division, Lawrence Berkeley National Laboratory, 
    1 Cyclotron Road MS~50B-4206, Berkeley, CA, 94611, USA
}
\altaffiltext{2}
{
    Physics Division, Lawrence Berkeley National Laboratory, 
    1 Cyclotron Road, Berkeley, CA, 94720, USA
}
\altaffiltext{3}
{
    Laboratoire de Physique Nucl\'eaire et des Hautes \'Energies,
    Universit\'e Pierre et Marie Curie Paris 6, Universit\'e Paris Diderot Paris 7, CNRS-IN2P3, 
    4 place Jussieu, 75252 Paris Cedex 05, France
}
\altaffiltext{4}
{
    Department of Physics, Yale University, 
    New Haven, CT, 06250-8121
}
\altaffiltext{5}
{
    Physikalisches Institut, Universit\"at Bonn,
    Nu\ss allee 12, 53115 Bonn, Germany
}
\altaffiltext{6}
{
    Department of Physics, University of California Berkeley,
    366 LeConte Hall MC 7300, Berkeley, CA, 94720-7300, USA
}
\altaffiltext{7}
{
    Universit\'e de Lyon, F-69622, Lyon, France; Universit\'e de Lyon 1, Villeurbanne; 
    CNRS/IN2P3, Institut de Physique Nucl\'eaire de Lyon.
}
\altaffiltext{8}
{
    Centre de Recherche Astronomique de Lyon, Universit\'e Lyon 1,
    9 Avenue Charles Andr\'e, 69561 Saint Genis Laval Cedex, France
}
\altaffiltext{9}
{
    Research School of Astronomy \& Astrophysics, 
    Mount Stromlo Observatory, 
    The Australian National University,
    Cotter Road, 
    Weston ACT 2611 Australia
}
\altaffiltext{10} 
{
    Skymapper Fellow
}
\altaffiltext{11}
{
    Centre de Physique des Particules de Marseille, 163, avenue de Luminy - Case 902 - 13288 Marseille Cedex 09, France
}
\altaffiltext{12}
{
    Tsinghua Center for Astrophysics, Tsinghua University, Beijing 100084, China 
}
\altaffiltext{13}
{
    Center for Cosmology \& Particle Physics,
    New York University,
    4 Washington Place, New York, NY, 10003, USA
}
\altaffiltext{14} 
{
    National Astronomical Observatories, Chinese Academy of Sciences, Beijing 100012, China
}
\altaffiltext{15} 
{
    Department of Physics and Astronomy, 
    University of Utah, 
    Salt Lake City, UT, 84112, USA
}
\altaffiltext{16} 
{
    Steward Observatory, 
    University of Arizona, 
    933 North Cherry Avenue, 
    Tucson, AZ, 85719, USA
}
\altaffiltext{*} 
{
    deceased
}

\begin{abstract}
We present convincing evidence of unburned carbon at photospheric
velocities in new observations of 5 Type~Ia supernovae (\sneia) obtained
by the Nearby Supernova Factory.  These SNe are identified by examining
346 spectra from 124 SNe obtained before $+2.5$~d relative to maximum.
Detections are based on the presence of relatively strong \ion{C}{2}~\wl
6580 absorption ``notches'' in multiple spectra of each SN, aided by
automated fitting with the \synapps\ code.  Four of the 5 SNe in
question are otherwise spectroscopically unremarkable, with ions and
ejection velocities typical of \sneia, but spectra of the fifth exhibits
high-velocity ($v > 20,\!000$~\kms) \ion{Si}{2} and \ion{Ca}{2}
features.  On the other hand, the light curve properties are
preferentially grouped, strongly suggesting a connection between
carbon-positivity and broad band light curve/color behavior: Three of
the 5 have relatively narrow light curves but also blue colors, and a
fourth may be a dust-reddened member of this family.  Accounting for
signal-to-noise and phase, we estimate that $22^{+10}_{-6}$\% of \sneia\
exhibit spectroscopic \ion{C}{2} signatures as late as $-5$~d with
respect to maximum.  We place these new objects in the context of
previously recognized carbon-positive \sneia, and consider reasonable
scenarios seeking to explain a physical connection between light curve
properties and the presence of photospheric carbon.  We also examine the
detailed evolution of the detected carbon signatures and the surrounding
wavelength regions to shed light on the distribution of carbon in the
ejecta.  Our ability to reconstruct the \ion{C}{2}~\wl 6580 feature in
detail under the assumption of purely spherical symmetry casts doubt on
a ``carbon blobs'' hypothesis, but does not rule out all asymmetric
models.  A low volume filling factor for carbon, combined with
line-of-sight effects, seems unlikely to explain the scarcity of
detected carbon in \sneia\ by itself.
\end{abstract}

\keywords{supernovae: general --- supernovae: individual
(SN~2005cf, SN~2005di, SN~2005el, SN~2005ki, SNF20080514-002)
\vspace{-0.30in}
}

\section{Introduction}

\setcounter{footnote}{0}

There is a well-established core population of Type~Ia supernovae
(\sneia) distinguished by striking spectroscopic homogeneity
\citep{branch1993, filippenko1997}.  The maximum light
ultraviolet-optical-infrared spectra of these SNe carry signatures of
predominantly intermediate-mass and iron-peak elements (often
identified: oxygen, magnesium, silicon, sulfur, calcium, and iron).  The
consistent presence of these features in the spectra of the \snia\ core
population contrasts with the intermittent detection of carbon lines.
Such carbon is detected in the earliest data, but fades away by the time
of peak brightness.

\sneia\ probably arise from the thermonuclear incineration of
carbon/oxygen white dwarfs in binary systems \citep[see][for a
review]{hillebrandt2000}.  Since carbon is consumed during the
explosion, constraints on its quantity, distribution, and incidence in
\snia\ ejecta help identify viable explosion mechanisms.  For example,
turbulent deflagration models predict that a large amount of unprocessed
carbon may be left over \citep[e.g.,][]{gamezo2003, ropke2005,
ropke2007} while models where the deflagration is followed by a
detonation phase can more completely burn it
\citep[e.g.,][]{hoflich2002, kasen2009}.  Interpreting oxygen signatures
is more ambiguous, since oxygen is produced as well as consumed in the
explosion.  Thus, constraints on carbon in \sneia\ are expected to be an
important piece of the explosion mechanism puzzle.

Unfortunately, multiple factors hinder an unbiased accounting of carbon
signatures in \snia\ spectra.  Challenges inherent in both discovering
and coordinating time-series spectroscopic follow-up of early, nearby
\sneia\ result in relatively poor statistics at phases optimal for
studying carbon (i.e., within 10 days after explosion or before about a
week prior to maximum light):  Early observations probe the outermost
layers of the SN, the region of the ejecta where carbon is most likely
to be found.  Optical carbon features are seldom conspicuous even at the
earliest phases, so noise may muddle detection in ``screening'' spectra
of young \sneia.  To achieve large numbers of SN candidates with a high
``purity'' of early \snia\ discoveries, SN surveys must cover as much
sky as possible while maintaining a few-day cadence; a difficult
endeavor.  Such surveys with associated on-demand access to spectroscopy
resources, like the Nearby Supernova Factory
\citep[SNfactory,][]{aldering2002} or the Palomar Transient Factory
\citep[PTF,][]{rau2009}, provide a direct way to address the relative
lack of spectroscopic time series sustained from the earliest phases.
Such data sets provide opportunities to obtain new constraints on
\snia\ physics.  By comparison, spectra obtained at and after maximum
light are already plentiful and well-studied.

The need to understand the \snia\ explosion mechanism and progenitor
system is as critical as ever, especially because of the important role
\sneia\ play in observational cosmology.  Observations of high-redshift
\sneia\ resulted in the discovery of Dark Energy, a revelation that has
changed our view of the Universe radically over the past decade.  The
Universe is not only expanding, but the rate of expansion itself is
\emph{accelerating}, in contrast to previous expectations that it would
be found to be decelerating \citep{riess1998, perlmutter1999,
astier2006, riess2007, mwv2007, kowalski2008, hicken2009a, sullivan2011,
suzuki2011}.  New techniques show promise for improving \snia\ distance
measurement precision \citep{bailey2009, mandel2009, wang2009a,
foley2011, blondin2011}.  However, the fact that \sneia\ couple Dark
Energy results to stellar evolution astrophysics amplifies concerns that
drift in emergent SN properties as a function of redshift could be
confused with a cosmology signal.  For example, the demographics of the
underlying progenitor population may drift over time, leading to a shift
in measured SN properties such as decline rate \citep{mannucci2005,
scannapieco2005, sullivan2006}.  Overcoming such problems requires
building a coherent picture of the progenitor system and explosion
mechanism.

In support of this effort, we consider the occurrence of unburned carbon
in a large set of early phase spectra obtained by the SNfactory.  Our
analysis supplements that of \citet{parrent2011} who conducted the first
comprehensive carbon signature survey using publicly available (and
some new) \snia\ spectroscopy.  Together, our studies find or verify
several ``definite'' or ``probable'' \ion{C}{2}~\wl 6580 detections in
\sneia\ that are neither super-Chandrasekhar candidates nor otherwise
obviously peculiar.  Six of 27 normal \sneia\ with SNfactory
observations before $-5$~d and with sufficient signal-to-noise (S/N)
have definite \ion{C}{2}~\wl 6580 detections.  Such normal SNe are the
foundation of the \snia\ Hubble diagram; studying observables correlated
with the incidence of carbon may provide new constraints on the
explosion mechanism, leading to new systematics controls or ruling out
the need for some.  In particular, we are interested in addressing the
question of whether carbon-positivity in \sneia\ is idiosyncratic
(distributed randomly within the population without regard to other
observables like light curve shape or color) or specifically correlated
with other parameters.  For example, if the frequency (or strength, or
persistence) of carbon signatures is somehow correlated with luminosity
or light-curve shape, explosion models need to reproduce this
relationship.  Our results are the strongest indication so far of 
such a connection between carbon-positivity and broad-band photometric
properties in normal \sneia.

The remainder of this article is arranged as follows.  In
\S~\ref{sec:selection}, the initial carbon detection procedure is
outlined and in \S~\ref{sec:data}, the selected sample is presented.
\S~\ref{sec:analysis} covers our analysis of the pre-maximum evolution
of the identified carbon features, overall characteristics of the
selected spectra, and light-curve/color properties of these SNe.
Details on individual objects are deferred to the Appendix.
\S~\ref{sec:discussion} brings together our results, provides the first
measurement of the \ion{C}{2}~\wl 6580 detection rate, and discusses our
findings in broader contexts.  In \S~\ref{sec:conclusion}, we draw
conclusions.

\section{Data Selection}

\label{sec:selection}

\begin{figure*}[ht]
    \centering
    \includegraphics[width=0.75\textwidth,clip=true]{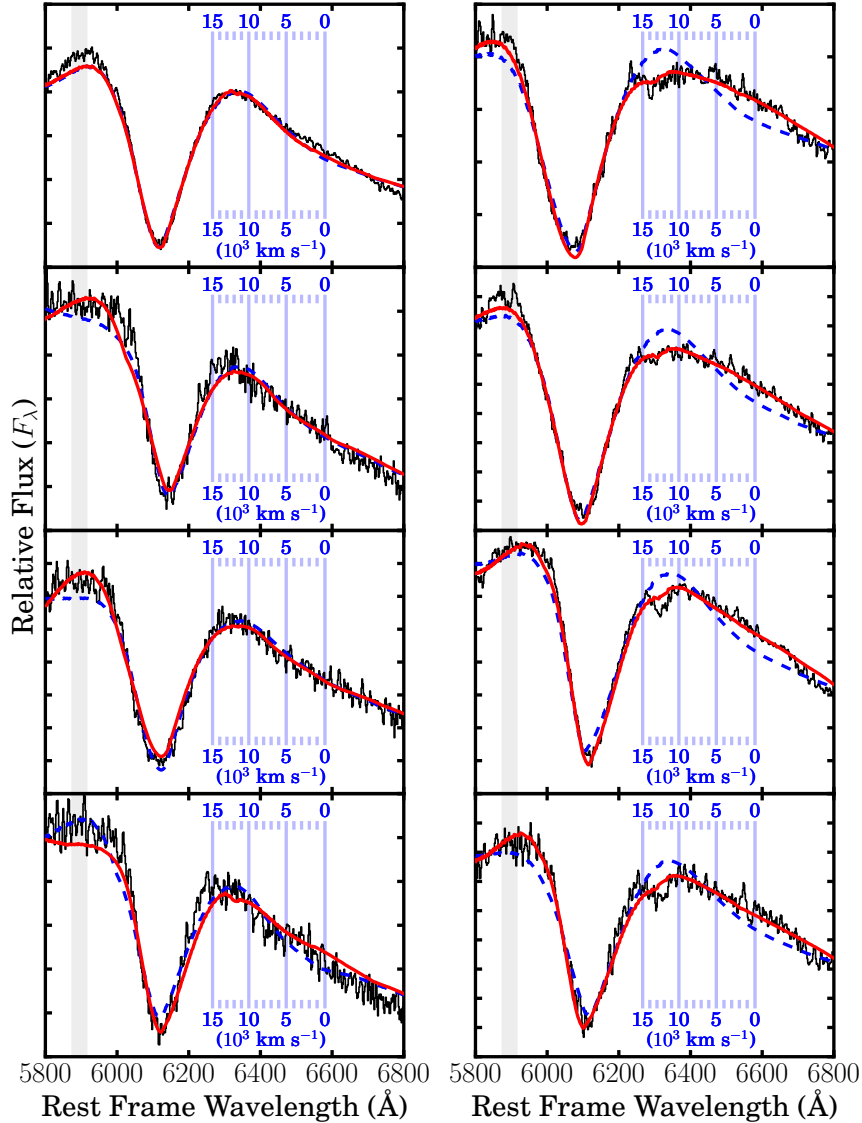}
    \caption{Identifying \ion{C}{2}~\wl 6580 features with the help of
    \synapps.  Each panel displays a set of fits to the \ion{Si}{2} \wl 6355
    region of each spectrum.  Respectively, red/solid and blue/dashed curves
    indicate fits without and with \ion{C}{2} opacity.  The left column shows
    non-detections, the right column shows \ion{C}{2}~\wl 6580 detection
    candidates.  Vertical grid lines represent blueshift relative to
    \ion{C}{2}~\wl 6580, and the shaded rectangle denotes the
    \ion{Na}{1}~D exclusion region.}
    \label{fig:feature}
\end{figure*}

The spectra used in this study were obtained by the SNfactory between
2004 and 2009 with the SuperNova Integral Field Spectrograph
\citep[SNIFS,][]{lantz2004}.  SNIFS is a fully integrated instrument
optimized for automated observation of point sources on a structured
background over the full ground-based optical
window at moderate spectral resolution.  It consists of a
high-throughput wide-band pure-lenslet integral field spectrograph
\citep[IFS, ``\`a la TIGER;''][]{bacon95,bacon00,bacon01}, a
multi-filter photometric channel to image the field in the vicinity of
the IFS for atmospheric transmission monitoring simultaneous with
spectroscopy, and an acquisition/guiding channel.  The IFS possesses a
fully-filled $6\farcs 4 \times 6\farcs 4$ spectroscopic field of view
subdivided into a grid of $15 \times 15$ spatial elements, a
dual-channel spectrograph covering 3200--5200~\AA\ and 5100--10000~\AA\
simultaneously, and an internal calibration unit (continuum and arc
lamps).  SNIFS is mounted on the south bent Cassegrain port of the
University of Hawaii 2.2~m telescope on Mauna Kea, and is operated
remotely.  Spectra of all targets were reduced using the SNfactory's
dedicated data reduction pipeline, similar to that presented in \S~4 of
\cite{bacon01}.  A key point is that all spectra analyzed and
presented have been corrected for telluric absorption.  A brief
discussion of the software pipeline is presented in \citet{aldering2006}
and is updated in \citet{scalzo2010}.  A detailed description of the
instrument and reduction pipeline will be published in the future.

The starting-point for the analysis is the sample of SNIFS spectra
obtained prior to $+2.5$~d with respect to rest-frame $B$-band maximum,
as determined by SALT2 light-curve fits \citep{guy2007} to \sneia\
observed on 5 or more nights.  This selection yields 346 epochs of
spectrophotometry from 124 individual \sneia.  We have excluded the
previously published carbon-positive SN~2006D \citep{thomas2007}, but
will include it in the discussion in \S~\ref{sec:discussion}.  The
sample includes spectroscopically ``normal'' \sneia\ \citep{branch1993},
overluminous SN~1991T-like and SN~1999aa-like \citep{li2001} \sneia, and
candidate super-Chandrasekhar-mass \sneia\ \citep{howell2006}.  A range
of light curve stretch values $0.65 < s < 1.16$ \citep{goldhaber2001} is
represented \citep[corresponding post-peak decline rate $1.82 > \Delta
m_{15}(B) > 0.73$,][]{guy2007}.  However, there are no subluminous
SN~1991bg-like events following the most strict definition \citep[i.e.,
exhibiting a prominent \ion{Ti}{2} absorption trough at maximum
light;][]{branch1993}.

To help identify candidate carbon features, we systematically
conduct fits to the relevant portion of each spectrum using the
\synapps\ code \citep{thomas2011}.  \synapps\ uses a highly
parameterized but fast SN spectrum synthesis technique, and as such does
not solve the full model SN atmospheres problem, in contrast to programs
like \phoenix\ \citep{hauschildt1997}, \cmfgen\ \citep{hillier1998}, or
\sedona\ \citep{kasen2006}.  The other codes mentioned are examples
designed for full/partial non-local thermodynamic equilibrium radiative
transfer, enabling detailed post-processing of stellar explosion models
for comparison to observations.  \synapps\ is useful in the opposite
sense, producing constraints on explosion models from observations,
since it identifies ion species and estimates their characteristic
ejection velocities.  Given the schematic radiative transfer model that
\synapps\ uses and its as yet unquantified systematic uncertainty, we
employ the fit results as an adjunct in our examination of the spectra
rather than as a direct quantitative discriminant.  The radiative
transfer assumptions underlying \synapps\ match those of the well-known
\synow\ SN spectrum synthesis program \citep{fisher2000, branch2009}:
Spherical symmetry, a sharply-defined blackbody-emitting
pseudo-photosphere, pure-resonance line transfer under Sobolev
formalism, with line opacity parameterized radially using convenient
functional forms and in wavelength assuming Boltzmann excitation.
\synapps\ combines this parameterized spectrum synthesis calculation
with a parallel non-linear optimization framework \citep{kolda2005,
gray2005, griffin2006} to reduce the need for tedious interactive
adjustment of fit parameters and to assure more systematic sampling of
the parameter space.  

We use \ion{C}{2}~\wl 6580 as our primary carbon indicator.  We consider
it the best option, for reasons we now describe.  \ion{C}{2}~\wl\wl 4745
and 7234 may also be useful \citep[and have been previously identified,
e.g.,][]{thomas2007} but are much weaker at the temperatures of
interest, and \wl 4745 is very likely to be blended with other lines.
Elsewhere, \ion{C}{2}~\wl 4267 has been considered as a carbon tracer
\citep[e.g.,][]{branch2003, howell2006, thomas2007} but it is
problematic: It does not seem to produce a reliable, distinct feature in
the spectrum, and is vastly weaker than \ion{C}{2}~\wl 6580 at the
relevant temperatures.  There is some evidence for the high-excitation
\ion{C}{3}~\wl 4649 line in the spectra of SN~1999aa-like
\citep{garavini2004} and SN~1991T-like \citep{parrent2011} \sneia, but
it is the only strong optical \ion{C}{3} feature and is not reliably
detected in the spectra of less luminous objects.  \ion{C}{1} lines are
mostly too red or too weak (e.g., \wl\wl9087, 9406, 10693) for us to
consider with SNIFS data \citep{marion2006, marion2009}.  Thus,
\ion{C}{2}~\wl 6580 seems to be the best choice, especially considering
the observational constraints.

As \citet{parrent2011} demonstrated (their Figure~9), \ion{C}{2}~\wl
6580 absorptions caused by carbon at higher velocities (much faster than
that at the photosphere at the phases of interest) will disappear into
the \ion{Si}{2}~\wl 6355 absorption trough where they cannot be detected
reliably.  Hence, inferences about carbon from the \ion{C}{2}~\wl 6580
notch are limited to lower velocities.  Fortunately, the low velocity
regime is what is of greatest interest for constraining the explosion
mechanism.  Carbon at the highest velocities may not be surprising; the
presence of carbon is much more variable at lower velocities in existing
models.

A final, but related, issue worthy of mention is the systematic error
that could be incurred when estimating the ejection velocity of carbon
directly from the position of the absorption notch instead of with the
aid of modeling, that of center-to-limb variation \citep{hoflich1990}.
Without the aid of any kind of spectral modeling, the velocity of the
carbon layer could be underestimated by up to a factor of 2.  However,
\synapps\ (at least) handles line blending explicitly, and the velocity
at the photosphere will be estimated by a fit to the full spectrum, so
we can be reasonably confident that the carbon is located near or just
above the photosphere.

The wavelength range selected for initial fitting extends from 5300 to
7300 \AA.  The pseudo-continuum level and line features are
simultaneously fit by \synapps, and all fits include lines from
\ion{O}{1}, \ion{Si}{2}, and \ion{S}{2}.  Lines from these ions are the
most important to consider in this wavelength region while the SN is on
the rise.  Strong iron-peak lines only become important after the phases
considered here.  A second-order polynomial warping function multiplied
against the synthetic spectrum is included to mitigate limitations of
the parameterized model continuum, which does not affect our ability to
detect a \ion{C}{2}~\wl 6580 absorption ``notch,'' a much smaller scale
feature.  Since \synapps\ models the SN spectrum but not the intervening
gas, we mask out rest-frame wavelengths between 5875 and 5915 \AA, a
generous allowance to exclude host galaxy interstellar (or
circumstellar) \ion{Na}{1}~D contamination, based on the width of the
strongest \ion{Na}{1}~D feature observed in the sample
\citep[SNF20080720-001,][]{atel1624}.  All spectra presented and fit by
\synapps\ are corrected for observer-frame Milky Way ($R_V = 3.1$)
extinction using the CCM \citep{cardelli1989} dust law, with color
excess estimates from \citet{schlegel1998}.

Three fits are performed to each spectrum to find candidate \ion{C}{2}
features.  The first set of fits including only \ion{O}{1}, \ion{Si}{2},
and \ion{S}{2} provides a carbon-free baseline.  The second set of fits
adds \ion{C}{2} with a ``photospheric'' Sobolev optical depth profile,
meaning that the minimum velocity of the profile is constrained to equal
the velocity at the photosphere.  The third set of fits substitutes a
``detached'' \ion{C}{2} optical depth profile, with the minimum velocity
of the profile unconstrained with respect to the photospheric velocity.
In all 3 fits, the Sobolev opacity declines with velocity exponentially.
A set of \ion{C}{2} fits allowing \ion{Si}{2} opacity to be detached was
also performed, but in practically all cases \synapps\ found a better
fit without \ion{Si}{2} detachment.

Some of the targets included in the analysis have previously been
identified or suspected of being carbon-positive, in particular
SNF20080514-002 \citep{atel1532}.  To reduce (somewhat) the risk of
confirmation bias, the process of organizing and executing the fits
includes concealing the identity of the parent SN of each spectrum and
its phase.  This is accomplished by shuffling the spectra and assigning
a unique integer identifier to each.  The mapping from spectrum to SN
and phase is kept in a hash file that is only consulted after the carbon
detections have been made.  This procedure, along with the \synapps\
fitting, forced us to consider more spectra more carefully.  It also
concealed the measured light curve fit parameters while identifying
strong carbon features, ensuring that any correlation between them and
the presence of carbon would also be uncovered without bias.

For a detection to be considered definitive, we require that \ion{C}{2}
be detected in at least 2 spectra of an object's time-series, either
from the same night of observation or across multiple consecutive
nights.  The final \ion{C}{2} identifications are confirmed by visual
inspection and comparison of the fits in the region of the \ion{Si}{2}
\wl 6355 emission peak.  Cases where the fits with and without
\ion{C}{2} are basically identical are considered non-detections (the
addition of \ion{C}{2} opacity does not improve the fit).  Fits with
\ion{C}{2} that differ from the baseline fit, and moreover improve upon
it, are flagged for closer scrutiny.  Several example fits are shown in
Figure~\ref{fig:feature}.  While the detailed agreement in all cases is
not perfect, the \synapps\ fits greatly improve our confidence in the
\ion{C}{2} identifications we otherwise would have made purely by visual
inspection. \synapps\ also highlighted cases we might have overlooked,
but in practice few of these cases ultimately made it into the sample,
usually because they fail the 2-spectrum cut or were less convincing.

\section{The Carbon-Positive \sneia}

\label{sec:data}

\begin{deluxetable*}{lllrrcl}
    \tabletypesize{\scriptsize}
    \tablecaption
    {
        New \ion{C}{2}-Positive \sneia.
    }
    \tablewidth{0pt}
    \tablehead
    {
        \colhead{Supernova\tablenotemark{a}}                & 
        \multicolumn{2}{c}{Host Galaxy\tablenotemark{b}}    &
        \multicolumn{2}{c}{SALT2 Fits}                      &
        \colhead{$B$-band Maximum}                          &
        \colhead{Rest Frame Phases}                         \\
        \colhead{}                                          &
        \colhead{Name}                                      &
        \colhead{$z_{helio}$}                               &
        \colhead{$x_1$}                                     &
        \colhead{$c$}                                       &
        \colhead{Date (UTC)}                                &
        \colhead{of Detection}
    }
    \startdata
    SN~2005di                   & MGC~-04-52-46  & $0.025298$ & $-1.35 \pm 0.22$ & $ 0.47 \pm 0.03$ & 2005-08-26.2 & $-11.3,-8.6$               \\
    SN~2005el                   & NGC~1819       & $0.014910$ & $-2.20 \pm 0.18$ & $-0.14 \pm 0.03$ & 2005-10-03.6 & $-6.9\ (\times 2)$         \\
    SN~2005ki\tablenotemark{c}  & NGC~3332       & $0.019207$ & $-1.89 \pm 0.08$ & $-0.01 \pm 0.01$ & 2005-12-01.2 & $-10.4,-8.4$               \\
    SNF20080514-002             & UGC~8472       & $0.022064$ & $-2.00 \pm 0.15$ & $-0.07 \pm 0.02$ & 2008-05-26.5 & $-10.0,-7.9,-5.0,-3.1?$    \\
    SN~2005cf                   & MCG~-01-39-003 & $0.006461$ & $-0.38 \pm 0.19$ & $+0.07 \pm 0.02$ & 2005-06-12.9 & $-9.4$
    \enddata
    \tablenotetext{a}{Discovery credits (all dates UTC). Lick Observatory Supernova Search \citep[LOSS,][]{filippenko2001}: SN~2005di on
    Aug.~12.4 \citep{cbet198}, SN~2005el on Sep.~19.5 \citep{cbet233}, SN~2005ki on Nov.~18.6 \citep{cbet294}, SN~2005cf on May~28.4
    \citep{cbet158}.  SNfactory: SNF20080514-002 on May~14.3 \citep{atel1532}.}
    \tablenotetext{b}{Redshift credits.  MGC~-04-52-46 \citep{mathewson1996}.  RC3.9 \citep{rc3.9}: NGC~1819, MCG~-01-39-003.  Sloan Digital Sky Survey DR7
    \citep{sdss7}: NGC~3332, UGC~8472.}
    \tablenotetext{c}{Light curve parameters from a fit to data from \citet{hicken2009b} and \citet{contreras2010}.}
    \label{tab:list}
\end{deluxetable*}

The fit survey yielded 6 of 124 \sneia\ with definitive \ion{C}{2}~\wl
6580 absorption notch detections.  In one of these cases, SN~2005cf,
there is only a single-spectrum detection, but this SN is difficult to
discard since the detection is corroborated by other external data sets.
Of the 124 \sneia\ in the sample, 4 are super-Chandrasekhar-mass \snia\
candidates \citep[including SN~2007if, the only clear \ion{C}{2}~\wl
6580 detection among the 4,][]{scalzo2010}.  Removing these leaves 5 out
of 120 carbon-positive \sneia:  SNe~2005di, 2005el, 2005ki,
SNF20080514-002, and SN~2005cf (a special case). For our dataset,
spectra having the requisite S/N generally come from SNe at the lowest
redshifts. Due to the resulting small volume sampled by untargeted
surveys out to such redshifts, the sample here is dominated by \sneia\
from targeted searches.  These new detections are the focus of our
detailed analysis in the rest of this article.  Basic target information
is summarized in Table~\ref{tab:list} and selected pre-maximum spectra
of each object, up to the point where the carbon signatures disappear,
are shown in Figure~\ref{fig:spectra}.

A few characteristics of the spectra presented in
Figure~\ref{fig:spectra} are worthy of specific mention.  The spectra
of SN~2005di are quite red, but there is evidence (discussed below) that
this is due to extrinsic dust extinction.  SN~2005el, SN~2005ki, and
SNF20080514-002 all possess what appear to be pronounced flux excess
blueward of about 3600~\AA.  The spectra of these 3 objects are also
dramatically similar in terms of detailed feature morphology.  SN~2005cf
clearly stands apart, with its more modest relative flux levels blueward
of 3600~\AA, but broader \ion{Ca}{2}~H\&K and \ion{Si}{2}~6150~\AA\
absorptions, and apparent high velocity (HV) \ion{Ca}{2} infrared
triplet.

\subsection{SN~2005di}

SNIFS observed SN~2005di on 10 occasions between Aug.~14 and
Sep.~22.\footnote{All dates discussed in this article are UTC.}  The
first 3 spectra of SN~2005di appear in Figure~\ref{fig:spectra}.  The
$-8.6$~d spectrum used here is a co-addition of 2 consecutive exposures
on the same night.  Close examination of the time series reveals the
presence of narrow rest-frame \ion{Na}{1}~D absorption (see
Figure~\ref{fig:carbon}a).  This and the relatively red appearance of
the spectral energy distribution lead us to conclude that SN~2005di is
heavily extinguished by line-of-sight dust in its host galaxy or
circumstellar environment.  Since SNIFS exposure times were generally
not adjusted to compensate for high extinction, the spectra of SN~2005di
have lower S/N than typical \sneia\ observed by SNIFS
at the same redshift.  Still, \ion{C}{2}~\wl 6580 is detected in the
first spectrum and possibly in the second.

Initial spectroscopic classifications of SN~2005di were published by 2
groups.  \citet{cbet200} reported that a spectrum obtained Aug.~16
resembles the spectrum of another \snia\ a few days past peak, and
\citet{atel581} state that their Aug.~14 spectrum suggests that
SN~2005di is ``young,'' but neither report discusses \ion{C}{2}.  The
SNIFS quick-reduction pipeline classification indicated that SN~2005di
was a \snia\ about 1 week before maximum, and highly extinguished with
an estimated $E(B-V) \sim 0.5$.  Revisiting our quick-reduction spectrum
we find that the \ion{C}{2}~\wl 6580 absorption notch could have been
identified, but its significance was not obvious at the time. 

SN~2005di was used in the SNfactory spectral flux ratio luminosity
indicator study \citep{bailey2009}.  That study found the ratio of
flux measured at 642~nm to that at 443~nm at maximum light
($\mathcal{R}_{642/443}$) to be a good predictor of relative luminosity.
SN~2005di is the object in that study with the largest value
of $\mathcal{R}_{642/443}$, owing primarily to its red color.  Any
possible effect of \ion{C}{2}~\wl 6580 emission near 642~nm either as a
possible source signal or noise in the relationship between
$\mathcal{R}_{642/443}$ and Hubble diagram residual was not discussed,
since at the time of publication the presence of \ion{C}{2} had not been
noticed.  We expect our \ion{C}{2}~\wl 6580 detection to have no
implication for the flux ratio result, since that study was restricted
to spectra obtained within 2.5~d of maximum, when the carbon signature
is gone.

\subsection{SN~2005el}

Nineteen SNIFS visits of SN~2005el were conducted between Sep.~26 and
Nov.~20.  Figure~\ref{fig:spectra} includes the spectra from the first 2
of these.  The $-6.9$~d observation is a co-addition of the 2
observations obtained that night.  In this case, \ion{C}{2}~\wl 6580 is
only positively detected by \synapps\ on the first date of observation.
However, it is clearly present in both high S/N spectra taken that night
and the co-add.

SN~2005el is a very well-observed SN \citep{mwv2008, hicken2009a,
contreras2010}.  Its spectroscopic normality has been remarked upon in
the literature previously \citep[e.g.][]{phillips2007, hoflich2010}.
Initial classification reports indicated that SN~2005el was discovered a
few to several days prior to maximum \citep{cbet235, cbet245} but until
now no remarks have been made in the literature regarding the presence
of \ion{C}{2} in the spectra.  These classification spectra were
obtained the same night as the first SNIFS observations.  The SNIFS
quick-reduction pipeline in use during the SN~2005el campaign failed to
properly correct for a cosmic-ray hit coincident with the \ion{C}{2}~\wl
6580 notch in one spatial element of the 2 spectra obtained on the first
night.  Subsequent improvements to the pipeline later revealed the clear
\ion{C}{2}~\wl 6580 notch in both spectra. 

\subsection{SN~2005ki}

SNIFS follow-up of SN~2005ki was triggered Nov.~20.  Six nights of
observations were obtained in all, ending Dec.~2, around the time of
maximum brightness.  Follow-up ended prematurely due to equipment
failure at the UH~2.2~m telescope that forced an extended offline period
for SNIFS. Therefore, we adopt the date of maximum and light-curve fit
parameters from a SALT2 fit to more complete photometry published by
\citet{hicken2009b} and \citet{contreras2010}.  The first 3 spectra
of SN~2005ki are presented in Figure~\ref{fig:spectra}.  Relatively
weak \ion{C}{2}~\wl 6580 absorption is detected in the first
spectrum, and very marginally in the second.  Returning to the first
2 SNIFS quick-reduction spectra of this object, the presence of
\ion{C}{2}~\wl 6580 now seems obvious in retrospect.  Our
classification announcement made no mention of it \citep{atel659}.

\subsection{SNF20080514-002}

Twenty nights of spectroscopy of SNF20080514-002 were obtained with
SNIFS, starting May~16 and ending Jul.~9.  The first 4 spectra of
SNF20080514-002 appear in Figure~\ref{fig:spectra}.  The $-7.9$~d and
$-5.0$~d spectra are co-additions of 3 consecutive observations on each
respective night.  \ion{C}{2}~\wl 6580 is clearly detected in the first
2 spectra of this object, and probably in the third.  Though the
emission peak visible at the position of the notch in the fourth
spectrum looks flattened, we hesitate to posit a flux depression due to
\ion{C}{2} opacity in this case.  At this point, \synapps\ is able to
convincingly fit the feature with or without a weak \ion{C}{2}~\wl 6580
line.

As initially reported \citep{atel1532}, the early spectra of
SNF20080514-002 are quite blue, indicative of a \snia\ at or before a
week prior to maximum light.  Both spectra exhibit absorption lines
attributed to \ion{C}{2}~\wl 6580, but \ion{C}{2}~\wl 7234 is not as
readily apparent as in, say, SN~2005el.  \ion{C}{2}~\wl 4745 was also
mentioned, but is a much weaker and more blended line than the others.

Like SN~2005di, SNF20080514-002 was used in the \citet{bailey2009} flux
ratio study.  In marked contrast to SN~2005di, this target has an
extremely \emph{blue} color but is well within the core of the Hubble
residual versus $\mathcal{R}_{642/443}$ distribution (their Figure~2).
As with SN~2005di, the use of the at-maximum spectrum for the flux ratio
study limits the influence of \ion{C}{2}~\wl 6580 on those results.

\subsection{SN~2005cf}

\begin{figure*}[htbp]
    \centering
    \includegraphics[width=0.90\textwidth,clip=true]{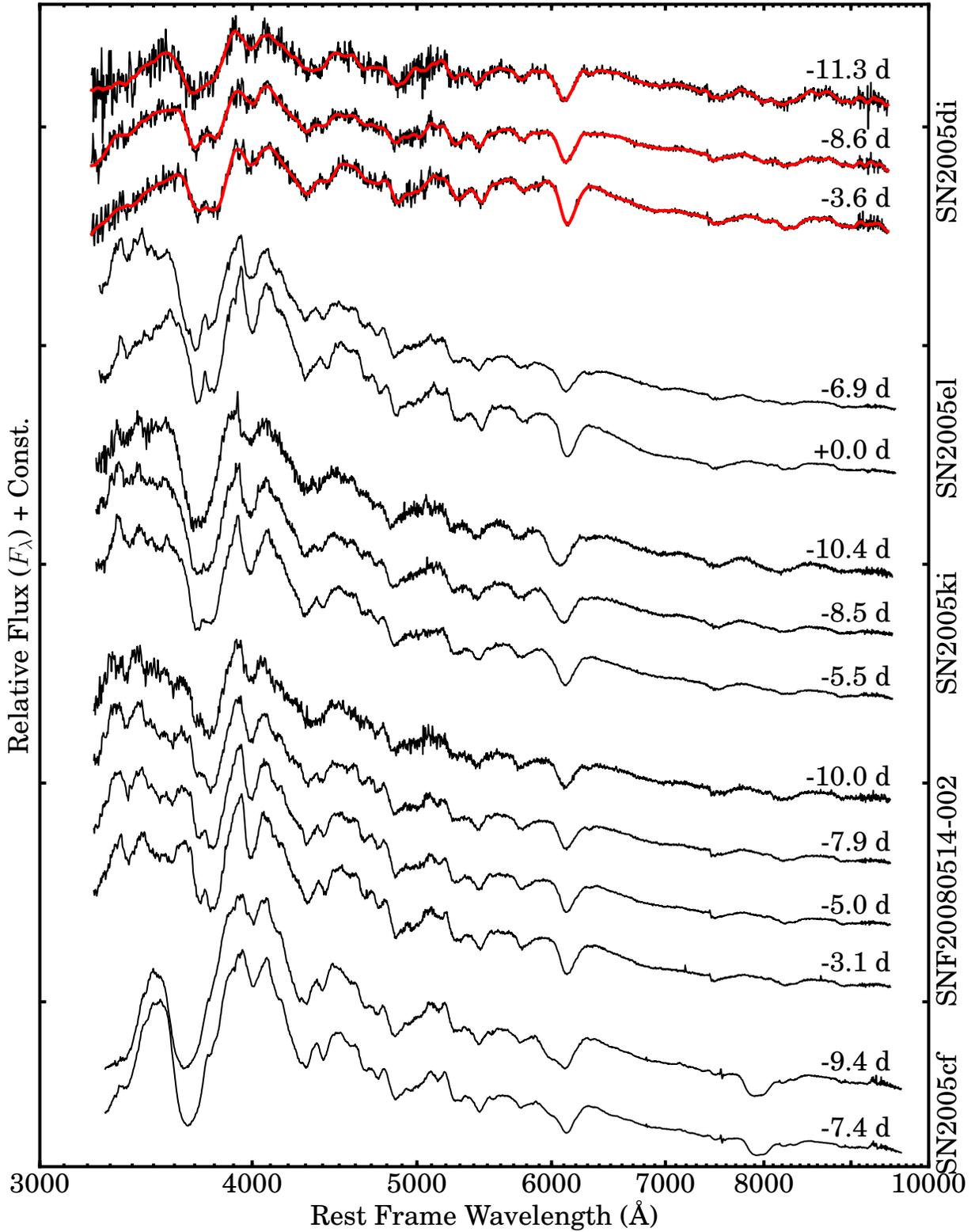}
    \caption{Pre-maximum SNIFS spectra of carbon-positive \sneia.
    Spectra up to the phase when \ion{C}{2}~\wl 6580 dissipates are
    included.  A smoothing filter \citep{savitzky1964} applied to the
    flux values is overlaid on the spectra of SN~2005di (in red) for
    presentation.  Phase labels denote rest-frame phase with respect to
    $B$-band maximum.  Spectra are labeled by SN name to the right.}
    \label{fig:spectra}
\end{figure*}

SNIFS observed SN~2005cf on 12 nights from Jun.~3 to Jul.~8.  The first
2 spectra are shown at the bottom of Figure~\ref{fig:spectra}.  A small
\ion{C}{2}~\wl 6580 absorption notch is visible in the $-9.4$~d spectrum
but becomes absent or nearly absent 2 days later.  Our selection
guidelines require detection in 2 distinct SNIFS spectra, so we should
technically disqualify SN~2005cf from further consideration.  However,
SN~2005cf is an object with extensive previously published spectroscopy
\citep{garavini2007, wang2009b}.  Neither previous spectroscopic study
nor the original classification announcement \citep{cbet160} discussed
the putative \ion{C}{2}~\wl 6580 notch visible in the spectra.  When
considering those observations, \citet{parrent2011} placed SN~2005cf
into their ``uncertain'' \ion{C}{2} detection category.  But given 3
independent data sets showing the same feature at roughly the same
phase, it now seems difficult to overlook the presence of \ion{C}{2} in
the case of SN~2005cf.  

\section{Analysis}

\label{sec:analysis}

\subsection{Spectroscopy}

\label{sec:carbon}

\begin{figure*}[htbp]
    \centering
    \begin{minipage}[]{0.48\textwidth}
        \includegraphics[width=\textwidth,clip=true]{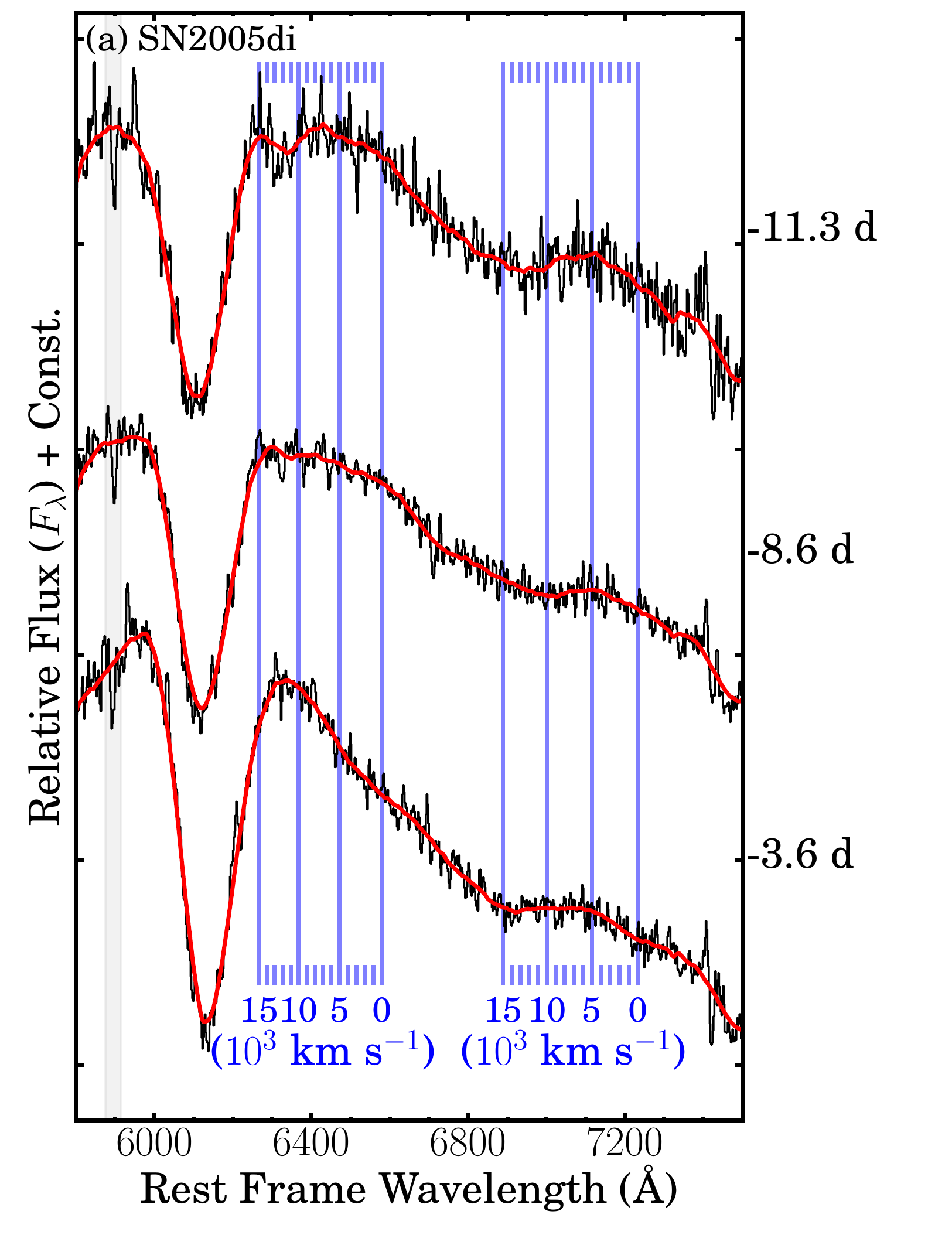}
    \end{minipage}
    \begin{minipage}[]{0.48\textwidth}
        \vspace{-1.0in}
        \includegraphics[width=\textwidth,clip=true]{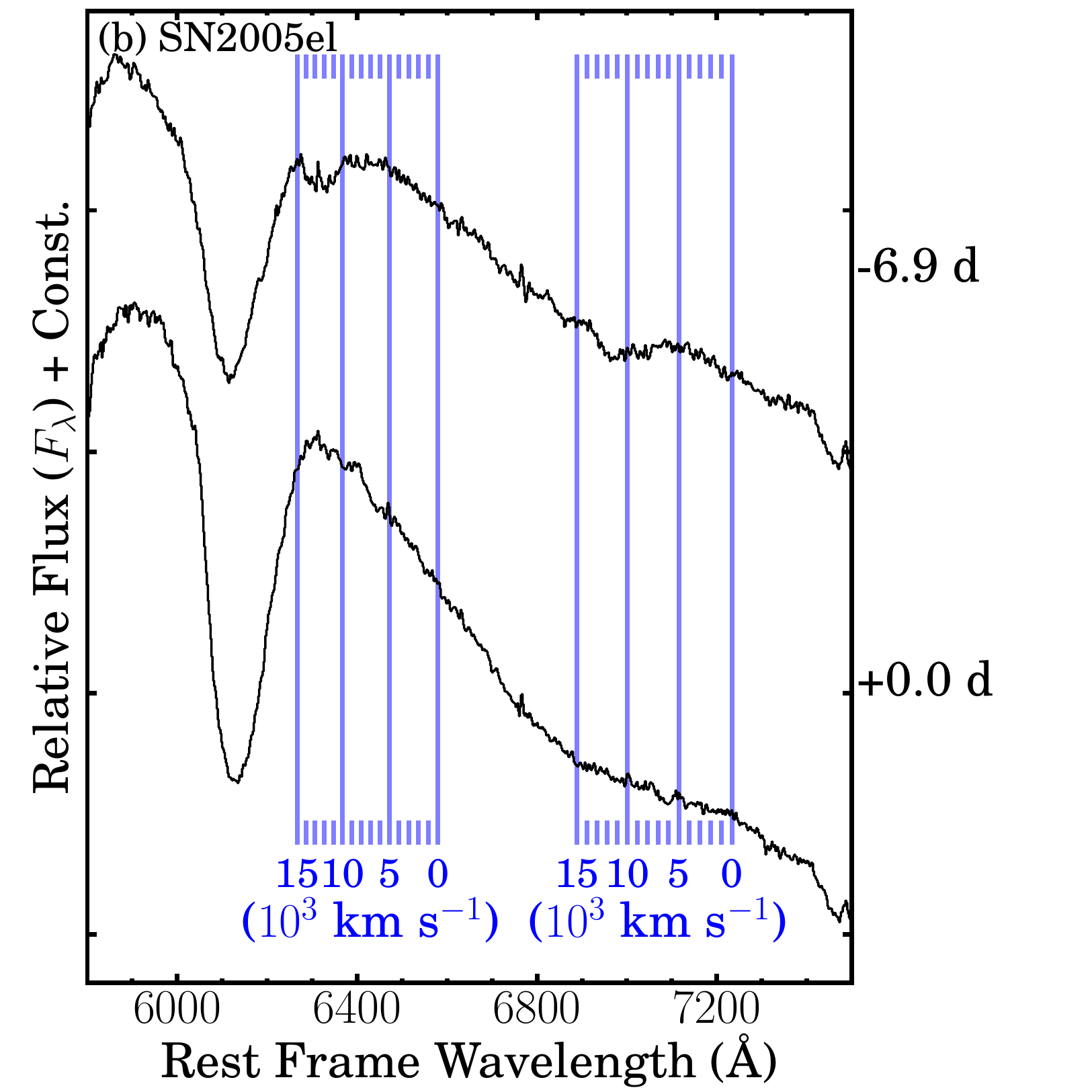}
    \end{minipage}
    \begin{minipage}[]{0.48\textwidth}
        \includegraphics[width=\textwidth,clip=true]{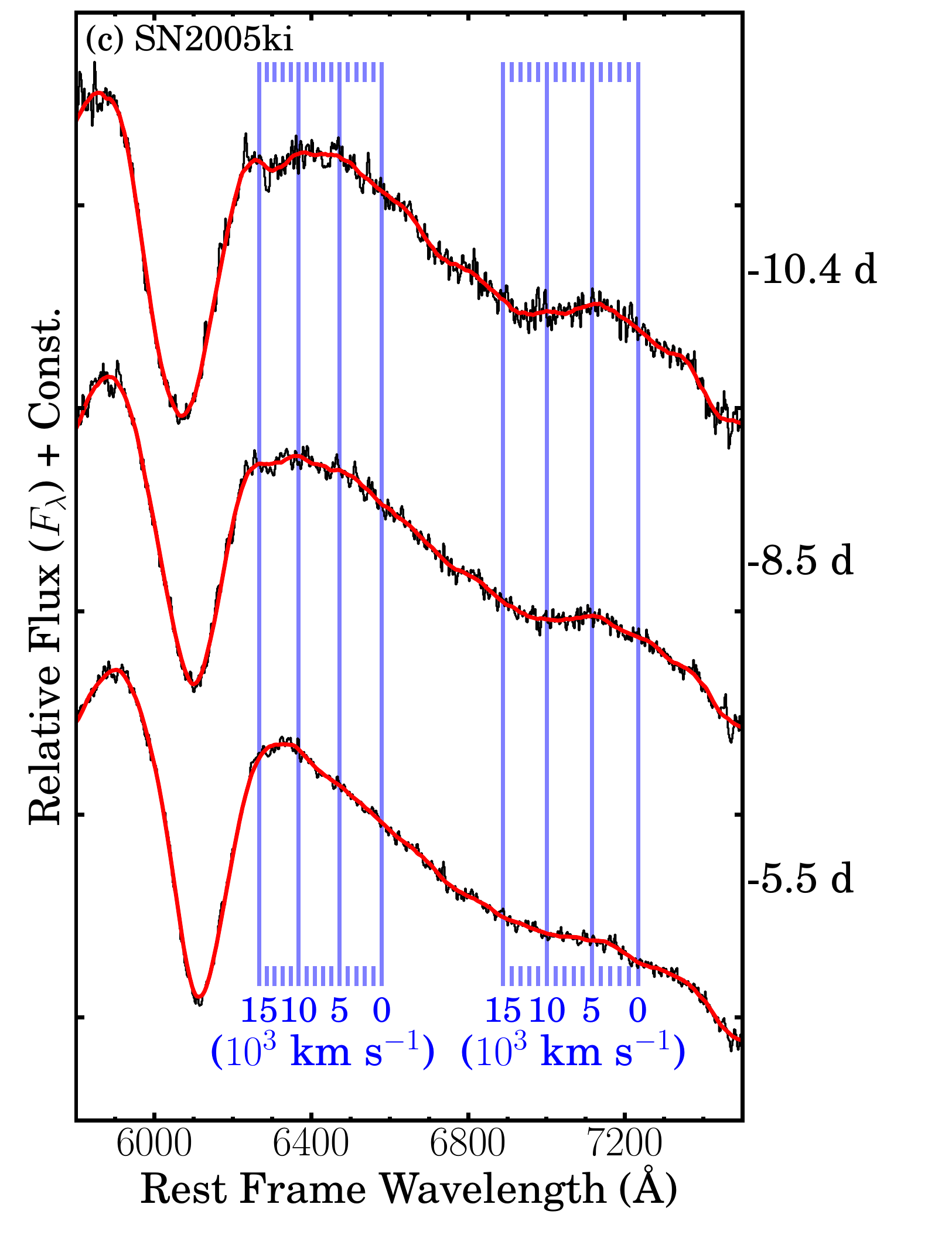}
    \end{minipage}
    \begin{minipage}[]{0.48\textwidth}
        \vspace{-0.9in}
        \includegraphics[width=\textwidth,clip=true]{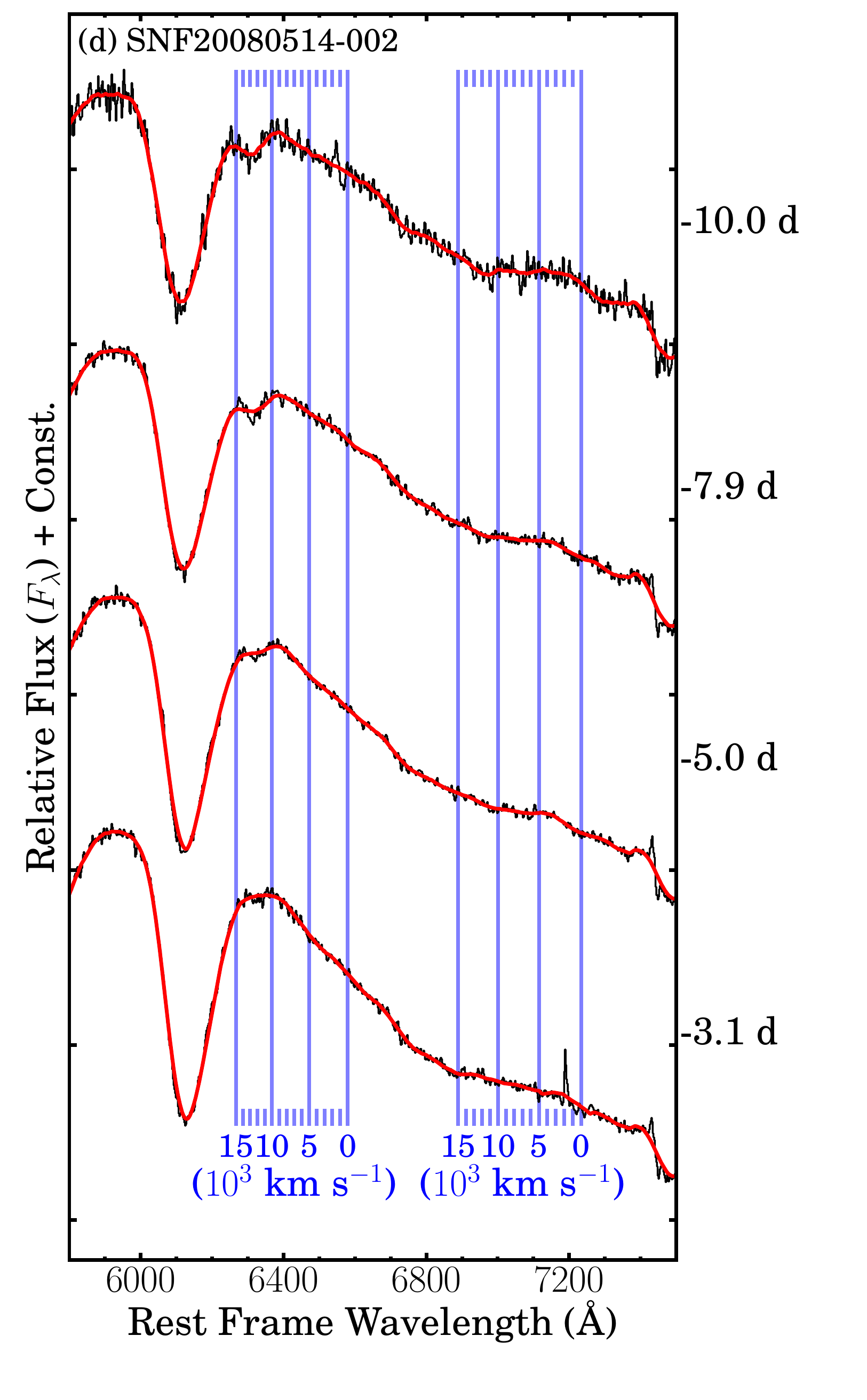}
    \end{minipage}
    \caption
    {
        Spectroscopic evolution of the 5800--7500~\AA\ region of the
        carbon-positive \sneia\ 2005di (a), 2005el (b), 2005ki (c),
        and SNF20080514-002 (d).  Smoothing-filtered flux values are
        overlaid (red line) for presentation purposes.  Blue vertical
        grid lines denote blueshift relative to 6580 and 7234 \AA, the
        rest wavelengths of the 2 strongest optical \ion{C}{2} lines.  The
        \ion{Na}{1}~D region is highlighted for SN~2005di, where narrow
        rest-frame line-of-sight absorption is clearly visible.
    }
    \label{fig:carbon}
\end{figure*}

\begin{figure}[htbp]
    \centering
    \includegraphics[width=0.48\textwidth,clip=true]{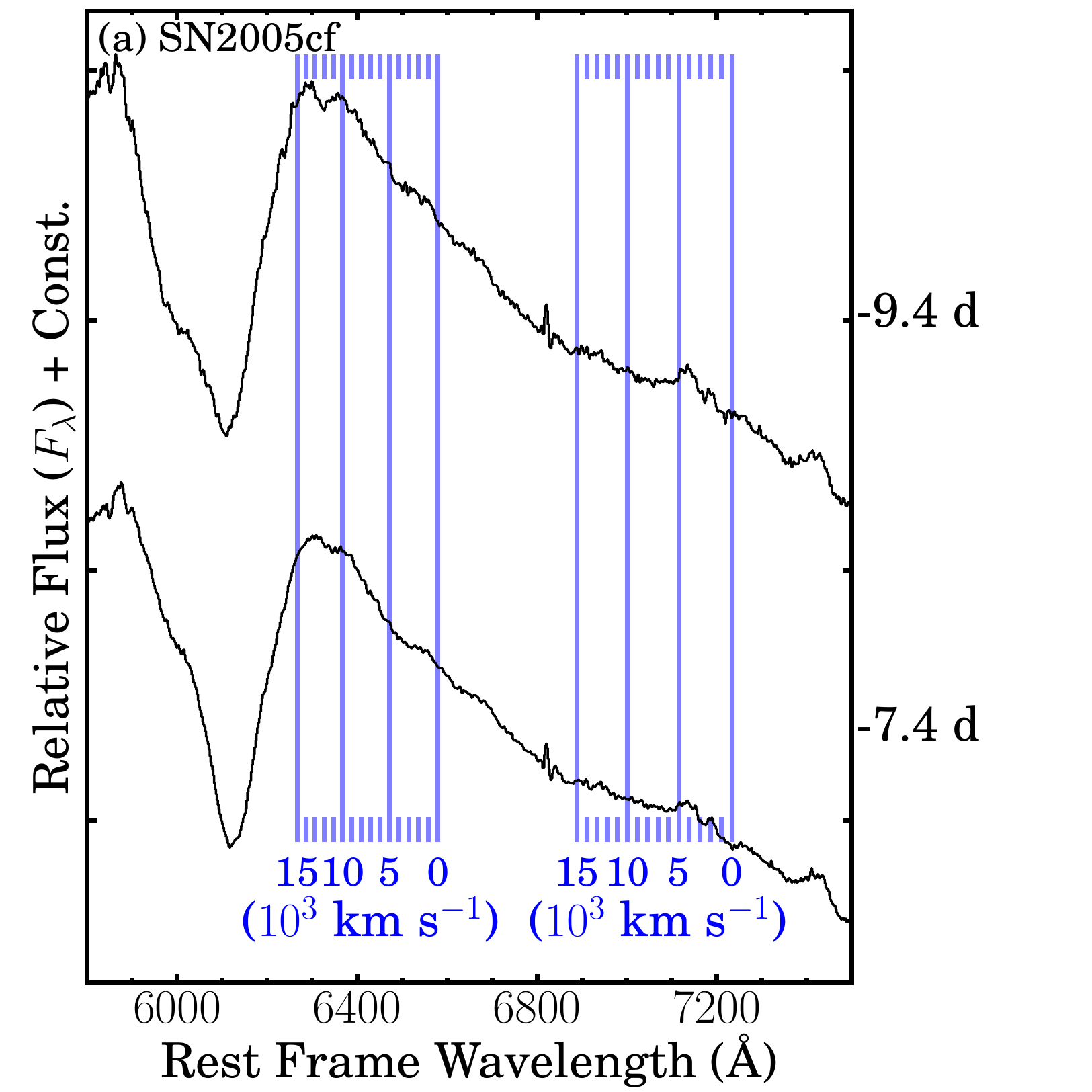}
    \includegraphics[width=0.48\textwidth,clip=true]{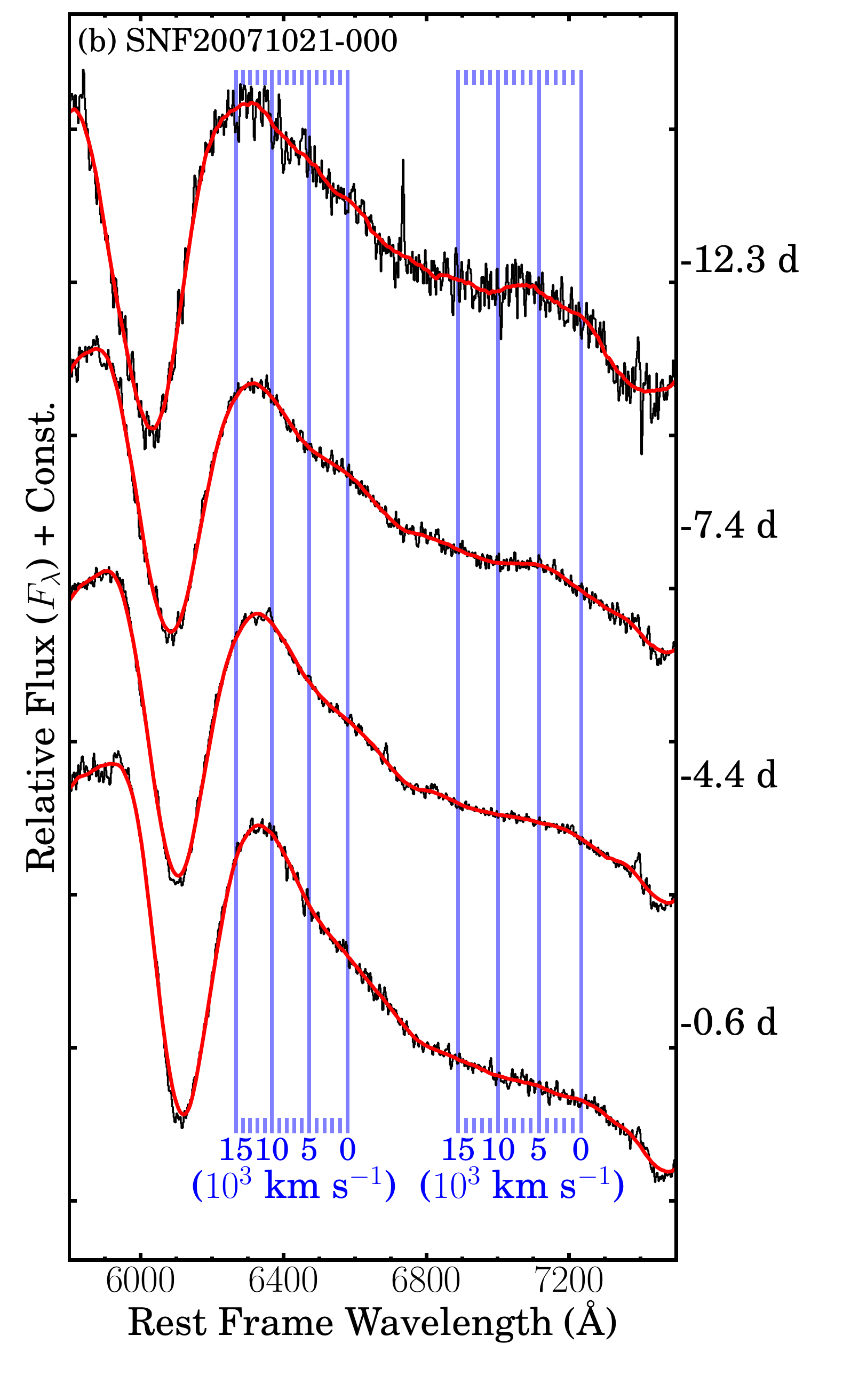}
    \caption
    {
        Probable \ion{C}{2} detection case SN~2005cf (a) and example
        non-detection case SNF20071021-000 (b).  Blue vertical grid
        lines denote blueshifts relative to \ion{C}{2} lines as in
        Figure~\ref{fig:carbon}.  In (b) the red overlaid curve is the
        smoothed flux.
    }
    \label{fig:others}
\end{figure}

The detailed evolution of the carbon features in the pre-maximum spectra
of SNe~2005di, 2005el, 2005ki, and SNF20080514-002 is depicted in
Figure~\ref{fig:carbon}.  SN~2005cf and a non-detection case for
comparison (SNF20071021-000) appear in Figure~\ref{fig:others}.  The
\ion{C}{2}~\wl 6580 notch is readily apparent, situated at or quite near
the rest-frame emission peak of the \ion{Si}{2}~\wl 6355 feature.  The
detailed evolution of the feature in each case is described in the
Appendix.

In the first 4 cases, the notch is strongest in the earliest spectrum.
It weakens gradually with time, and without appreciable wavelength
shift.  Generally, the notch is centered on an apparent blueshift of
$12,\!000$~\kms\ relative to rest-frame \ion{C}{2}~\wl 6580 (also
roughly the same velocity as observed in SN~2006D).  In SN~2005cf, the
notch disappears between the first and second spectrum.

Excluding SN~2005cf, we see that while the notch dissipates, the shape
of the spectrum just redward evolves as well.  The slope of the decline
in flux on the red side of the \ion{Si}{2}~\wl 6355 emission feature
becomes steeper while the notch weakens and the underlying
\ion{Si}{2}~\wl 6355 emission peak rises.  In some cases (SNe~2005di,
2005ki, and SNF20080514-002) the earlier spectra may even exhibit a mild
``flux plateau'' morphology.  Along with an absorption notch, such an
emission plateau could be a signature of a ``detached'' distribution of
\ion{C}{2} line opacity in a shell separated somewhat from the
photosphere \citep{jeffery1990}.  However, under spherical symmetry,
such a plateau should be roughly symmetric about the line rest
wavelength.  Visually, this does not seem to be the case here.

The notch in the spectrum of SN~2005cf is morphologically distinct from
the other cases, and appears to span a slightly smaller range in
blueshift ($10,\!000$ to $13,\!000$~\kms).  When the notch disappears,
there is at most only a small amount of evolution in the flux slope just
redward.

Confident identification of \ion{C}{2}~\wl 7234 is problematic.  In
almost none of the cases presented is there a clear absorption feature
at the same blueshift as the \ion{C}{2}~\wl 6580 notch.  At most, there
is only an inflection visible around 6900--7000~\AA\ and a slight
emission ``bump'' around 7100~\AA.  Visual inspection alone would seem
to be insufficient to draw more firm conclusions about the presence or
absence of \ion{C}{2}~\wl 7234 in general.

\begin{figure*}[htbp]
    \centering
    \begin{minipage}[]{\textwidth}
        \includegraphics[width=\textwidth,clip=true]{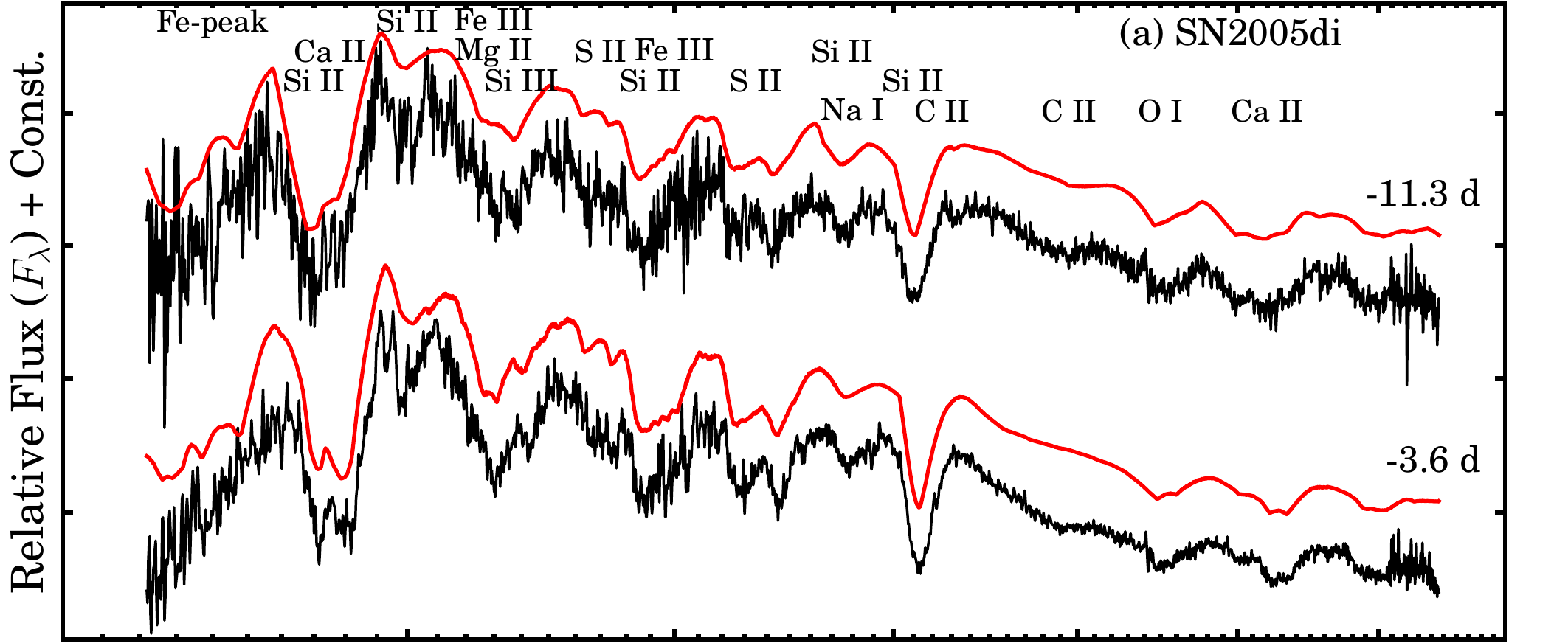}
    \end{minipage}
    \begin{minipage}[]{\textwidth}
        \includegraphics[width=\textwidth,clip=true]{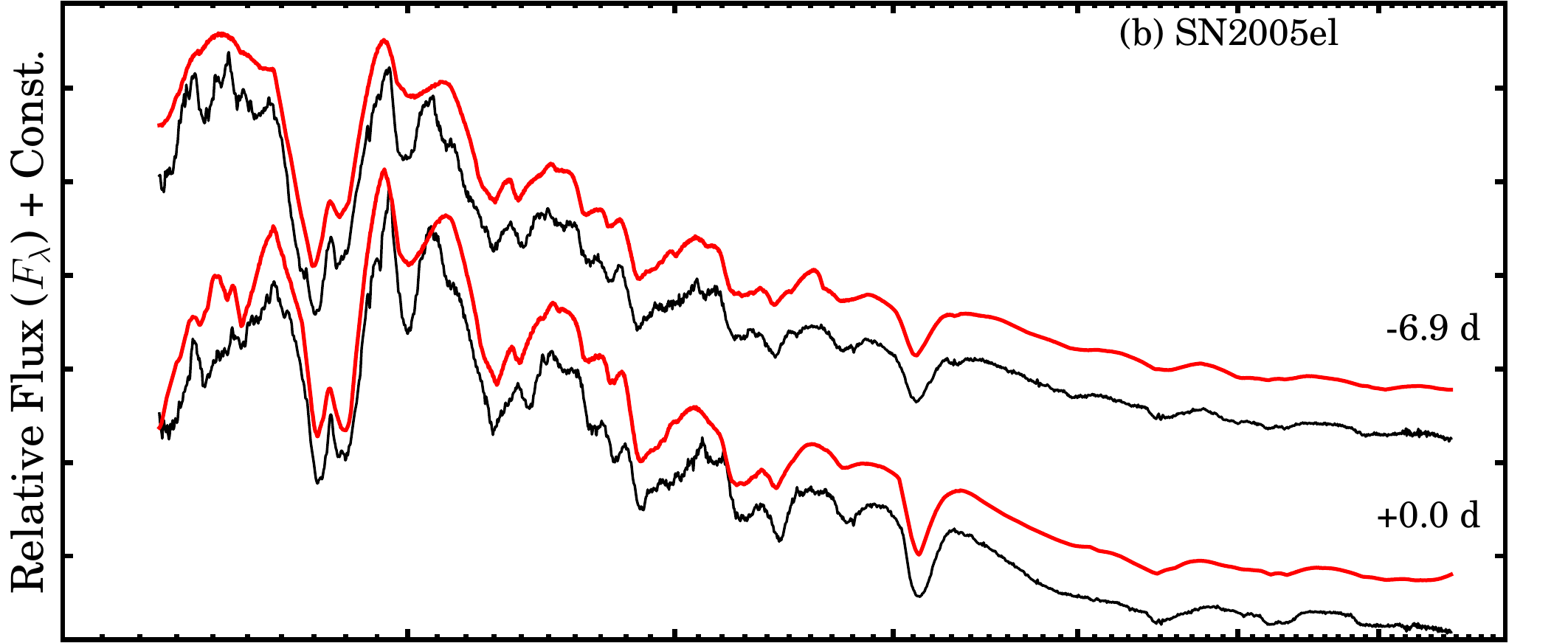}
    \end{minipage}
    \begin{minipage}[]{\textwidth}
        \includegraphics[width=\textwidth,clip=true]{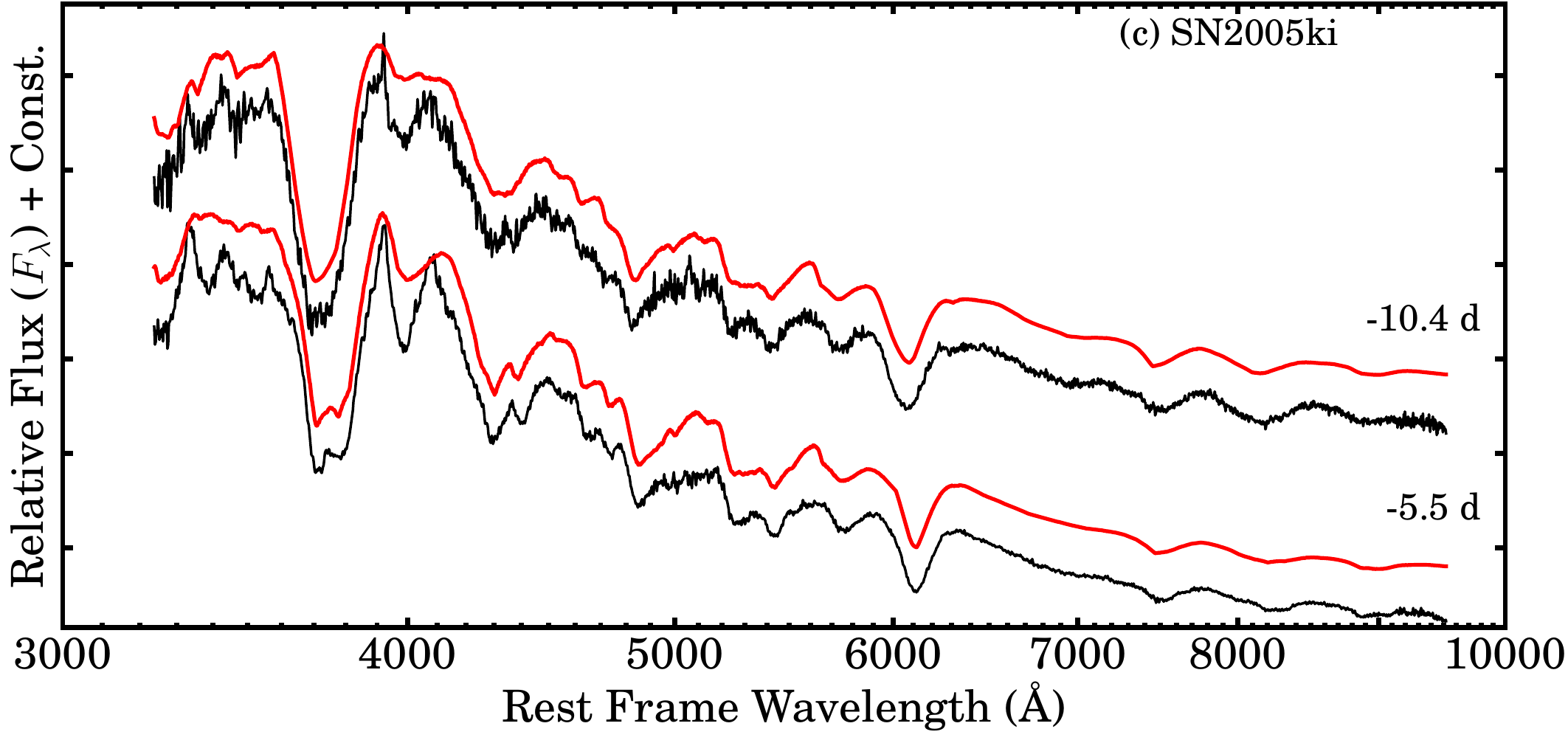}
    \end{minipage}
    \caption
    {
        \synapps\ fits (red lines) to selected pre-maximum spectra of
        SNe~2005di, 2005el, and 2005ki (panels a, b, and c
        respectively).  The fit to the first available spectrum of each
        SN is shown, followed by the first fit where \ion{C}{2} opacity
        is no longer invoked.  Major contributions to the fit are
        labeled by ion in panel (a).  Singly ionized atoms of Ti through
        Co contribute to the spectrum blueward of 3500~\AA.
    }
    \label{fig:synapps_1}
\end{figure*}

\begin{figure*}[h]
    \centering
    \begin{minipage}[]{\textwidth}
        \includegraphics[width=\textwidth,clip=true]{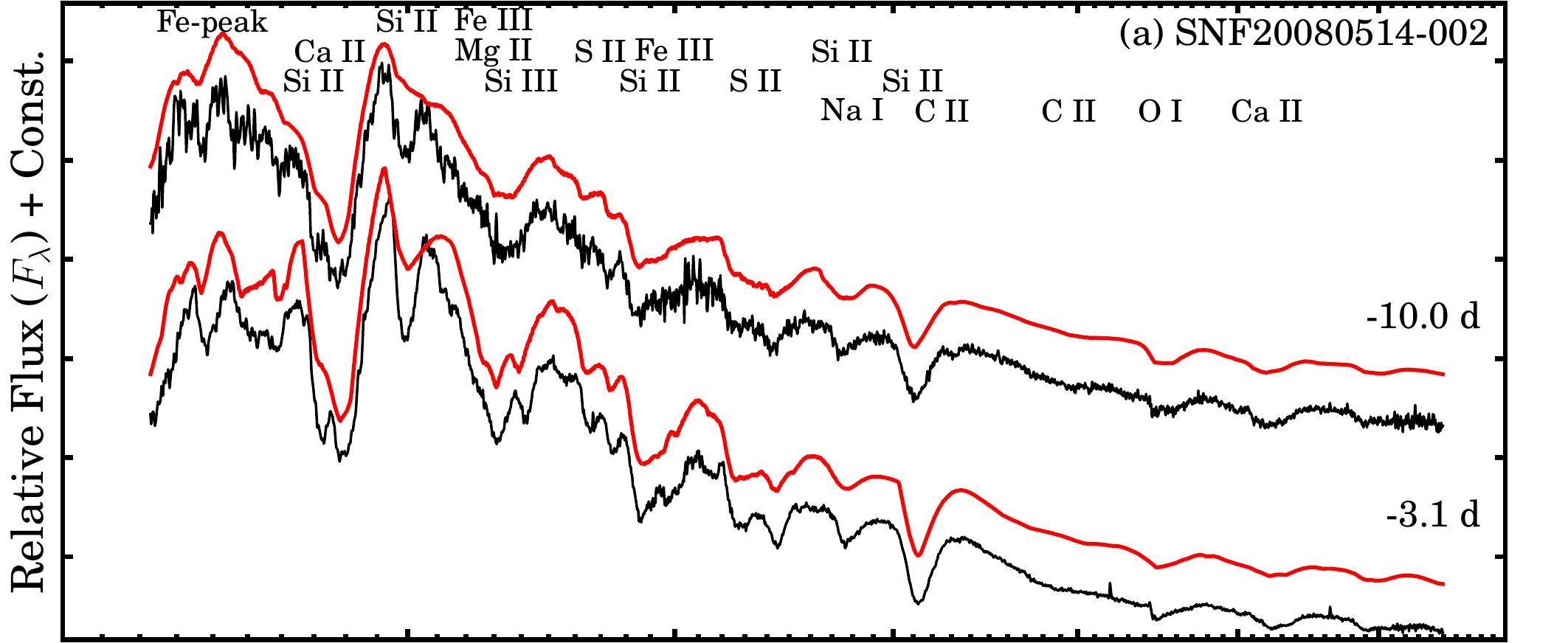}
    \end{minipage}
    \begin{minipage}[]{\textwidth}
        \includegraphics[width=\textwidth,clip=true]{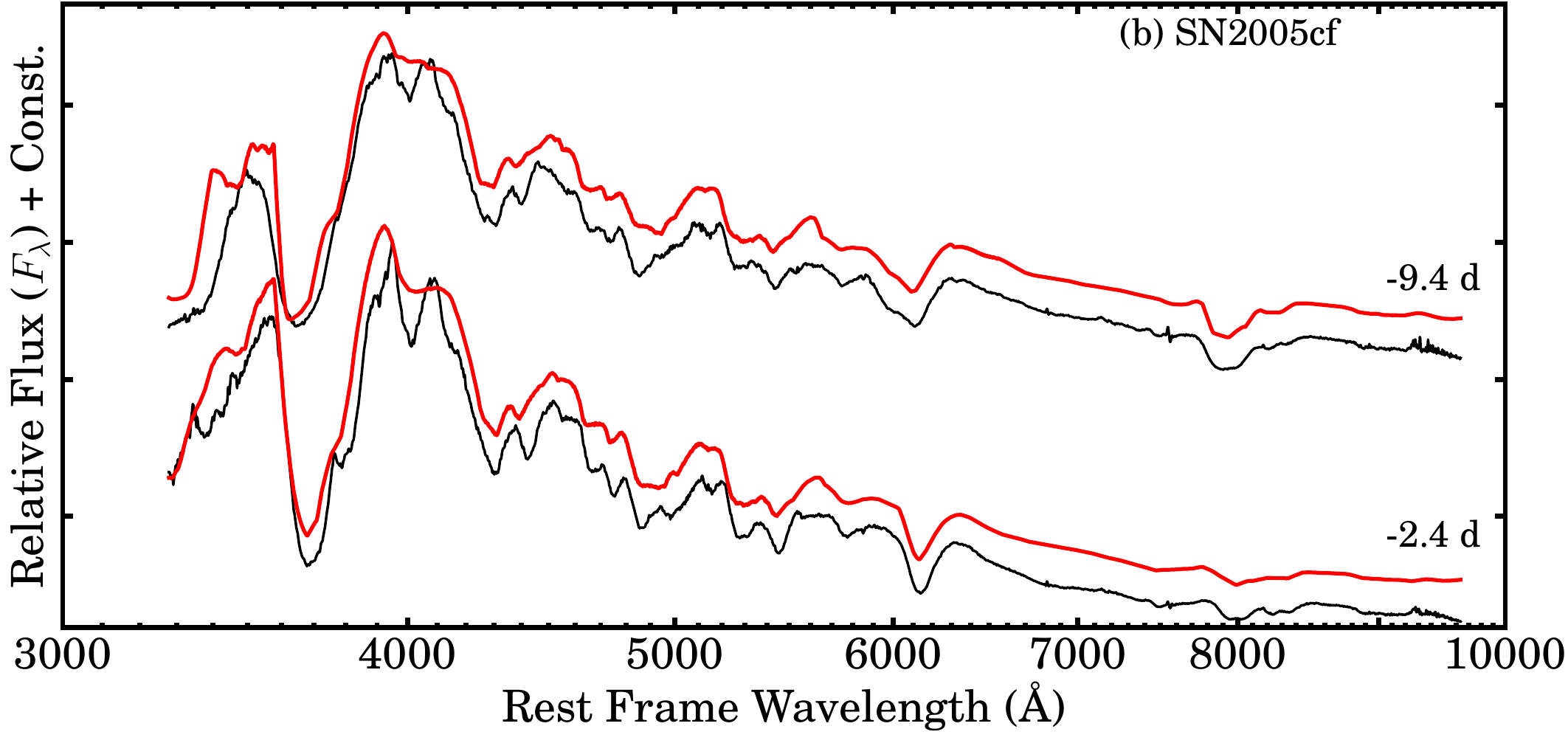}
    \end{minipage}
    \caption
    {
        \synapps\ fits (red lines) to selected pre-maximum spectra of
        SNF20080514-002 (a) and SN~2005cf (b).  The fit to the first
        available spectrum of each SN is shown, followed by the first
        fit where \ion{C}{2} opacity is no longer invoked.  Major
        contributions to the fit are labeled by ion in panel (a).
        Singly ionized atoms of Ti through Co contribute to the spectrum
        blueward of 3500~\AA.
    }
    \label{fig:synapps_2}
\end{figure*}

\begin{figure}[htbp]
    \centering
    \includegraphics[width=0.48\textwidth,clip=true]{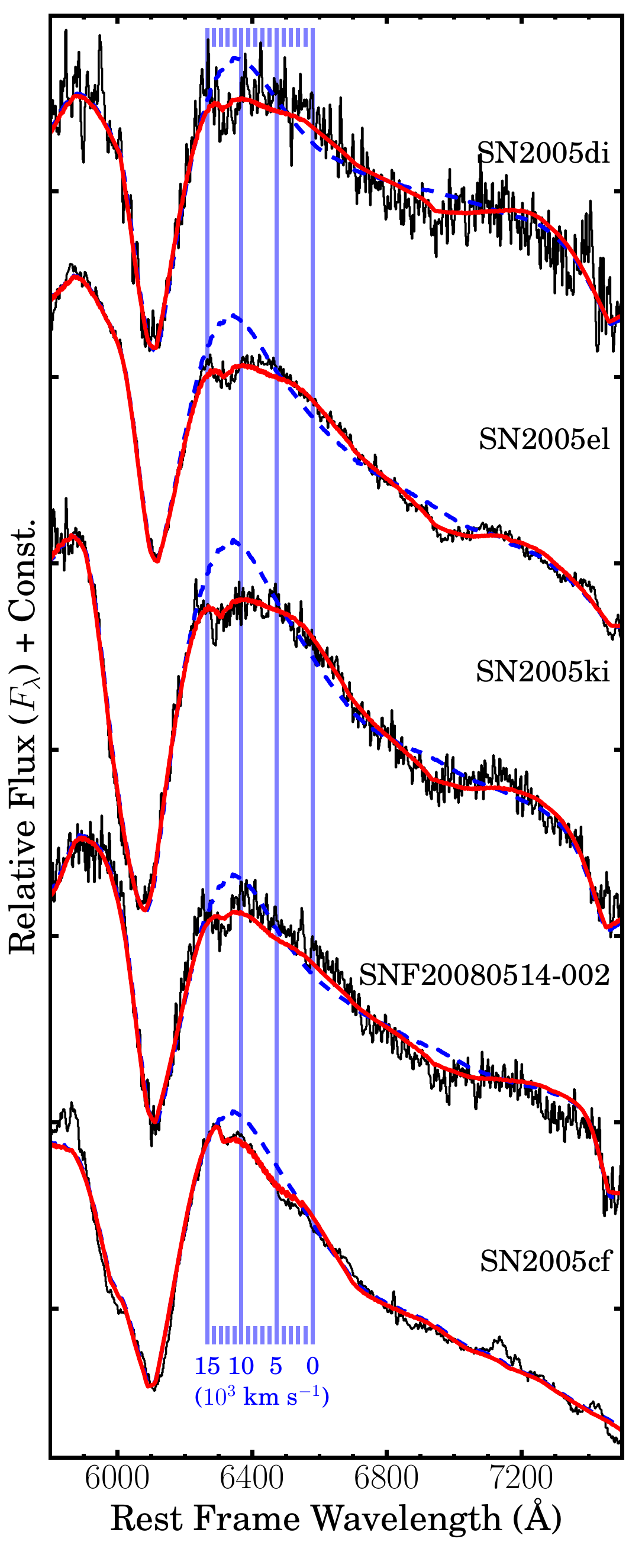}
    \caption
    {
        Action of \ion{C}{2} opacity in converged \synapps\ fits to the
        earliest observed spectrum of each SN.  The red/solid curve
        plots the synthesized spectrum from the best fit with all ions.
        The blue/dashed curve results when the final synthetic spectrum
        is recomputed without \ion{C}{2}.  The carbon opacity screens
        the emission peak from \ion{Si}{2}, in addition to creating the
        notch shape.
    }
    \label{fig:closeup}
\end{figure}

To examine the extent to which the selected targets are
spectroscopically normal \sneia, we again use \synapps, this time to fit
the full wavelength range of each pre-maximum spectrum of each selected
SN (detailed commentary is provided in the Appendix).  Close examination
of the \ion{C}{2} features in the full fits may also enable some
explosion-model independent statements about the distribution of carbon
in the SN ejecta.  Two selected fits for each SN appear in
Figures~\ref{fig:synapps_1} and \ref{fig:synapps_2}: One fit to the
first available spectrum, and another fit to the first spectrum where
\ion{C}{2} is absent.  In general the opacity profiles for \ion{C}{2}
lines decrease with time as expected.  Wavelengths of major ion
contributors to the spectra are marked at the top of either figure.
Except for the presence of \ion{C}{2}, the spectra of the first 4 
carbon-positive \sneia\ are anything but extraordinary --- typical
\snia\ ions at normal pre-maximum ejection velocities.  SN~2005cf is
another matter, since HV \ion{Si}{2} and \ion{Ca}{2} components are
mandatory for a good fit, as discussed previously \citep{garavini2007,
wang2009b}.

In each case, we find compelling evidence for opacity from \ion{C}{2},
\ion{O}{1}, \ion{Mg}{2}, \ion{Si}{2} and \ion{Si}{3}, \ion{S}{2},
\ion{Ca}{2}, singly-ionized ions of Ti through Co, and \ion{Fe}{3}.  For
both \ion{Si}{2} and \ion{Ca}{2} we enable 2-component fits: One for a
photospheric (undetached) component, and another for a HV (detached)
component.  The opacity profiles of all other ions are undetached from
the photosphere.  Optimal values for the velocity at the photosphere are
found in the range $12,\!000$ to $13,\!000$~\kms.  Including some
\ion{Na}{1} opacity does seem to improve the fit morphology around the
\ion{Si}{2} 5800~\AA\ absorption feature --- however this identification is
not unprecedented \citep[e.g.,][]{branch2005}.  In all cases, we were
able to reconstruct the basic structure redward of 3500~\AA\ quite well,
and at times had reasonable success fitting the ultraviolet (UV, where
lines from iron-peak elements dominate).  We do not expect exceptional
agreement in the UV, as the underlying assumptions of \synapps\ (in
particular the treatment of line opacity) begin to break down here.  In
general, the strengths of the features evolve smoothly with time, and no
abrupt changes are observed as the \ion{C}{2} becomes optically thin.

\synapps\ accounts for the \ion{C}{2}~\wl 6580 feature, including the
region just redward of the notch, rather well (see
Figure~\ref{fig:closeup}, red curve).  Variations with and without
detached \ion{C}{2} were tried, but again \synapps\ tended to set the
lower velocity of the \ion{C}{2} region to the velocity at the
photosphere, making the need for detachment uncompelling.  Analysis of
the fit residuals indicates that the \synapps\ model has insufficient
fidelity for achieving a reduced $\chi^2$ of 1, but we find no strong
indications of persistent residual structure.  

In Figure~\ref{fig:closeup}, the 5800--7500~\AA\ region of the first
spectrum of each carbon-positive SN is overlaid with the converged,
all-ions, full-wavelength \synapps\ fits (shown in red).  The blue curve
results when the fully converged synthetic spectrum is recomputed with
the \ion{C}{2} opacity deactivated, while all other settings retain
their converged values.  The blue curves are thus \emph{not} re-fits
without \ion{C}{2} included, in contrast to the example plots shown in
Figure~\ref{fig:feature}.  Thus, the change from the blue curve to the
red one illustrates what role \ion{C}{2} plays in the formation of the
spectrum quite clearly.  Not only is \ion{C}{2}~\wl 6580 responsible for
the absorption notch, it ``screens'' emission from bluer \ion{Si}{2}~\wl
6355 effectively enough that some emission relative to the
\ion{C}{2}-free spectrum (centered around 6600~\AA) is visible in the
top 4 cases.  Finally, Figure~\ref{fig:closeup} indicates that the flux
depression around 6900--7000~\AA\ may indeed be influenced by
\ion{C}{2}~\wl 7234 to some extent.  

\subsection{Light Curve and Color Properties}

\label{sec:lc}

\begin{figure*}[htbp]
    \centering
    \includegraphics[width=0.99\textwidth,clip=true]{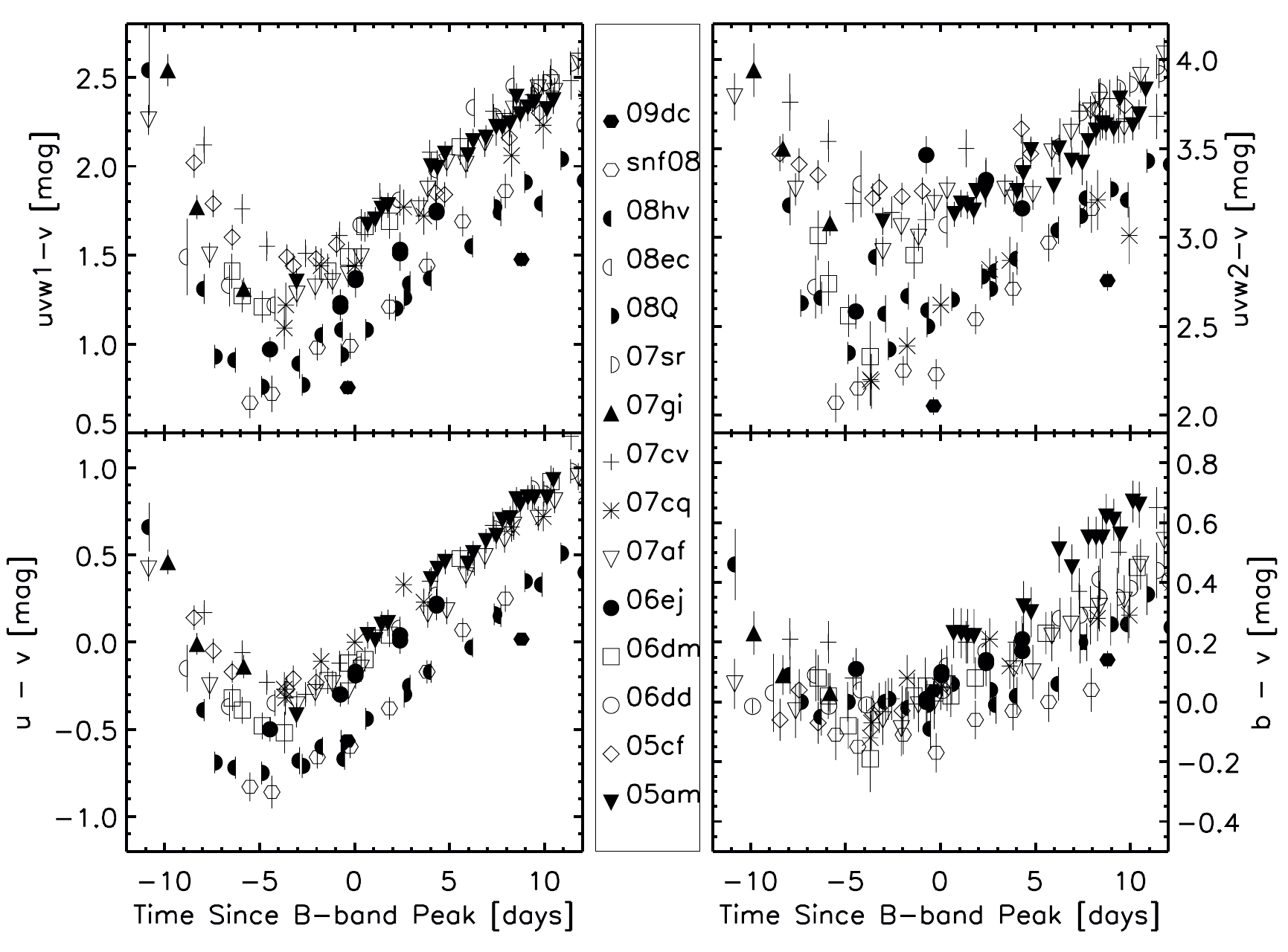}
    \caption{\snia\ colors from Swift/UVOT.  Note that SNF20080514-002
    (open hexagon), SN~2008Q (right filled semicircle), and SN~2008hv
    (left filled semicircle) are among the objects with the largest UV
    excess observed by Swift.  In particular, the objects are very
    clearly separated in $u-v$ color space from the rest of the sample.}
    \label{fig:uvot}
\end{figure*}

\begin{figure*}[ht]
    \centering
    \includegraphics[width=0.90\textwidth,clip=true]{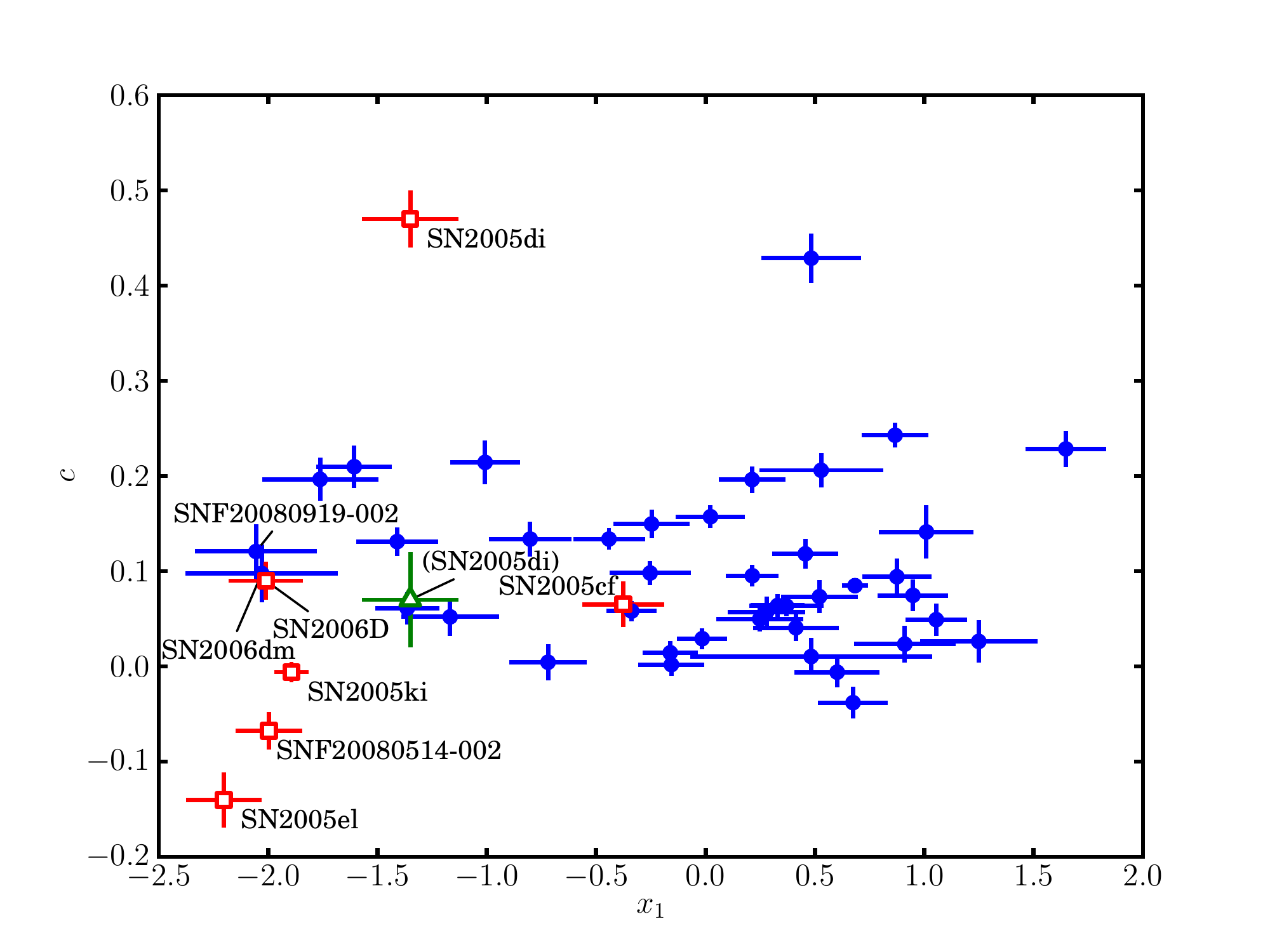}
    \caption
    {
        SALT2 fit parameters for \sneia\ followed with SNIFS having
        observations earlier than $-5$~d with respect to maximum light
        (narrower \snia\ light curves correspond to negative $x_1$,
        broader ones to positive $x_1$).  Red open squares mark the
        positions of carbon-positive \sneia, blue circles mark the
        others.  The green open triangle represents SN~2005di after
        removing a differential $E(B-V)$ obtained by warping the
        spectral energy distribution to match those of SN~2005el,
        2005ki, and SNF20080514-002.  Three of the 5 identified
        carbon-positive \sneia\ form a family of
        fast-decliner/blue-color objects.  Other \sneia\ (labeled by
        name, but blue circles) with similar show very marginal evidence
        for \ion{C}{2}~\wl 6580, but their first phase of observation is
        around $-7$~d.
    }
    \label{fig:lc_params}
\end{figure*}

\begin{figure*}[ht]
    \centering
    \includegraphics[width=0.90\textwidth,clip=true]{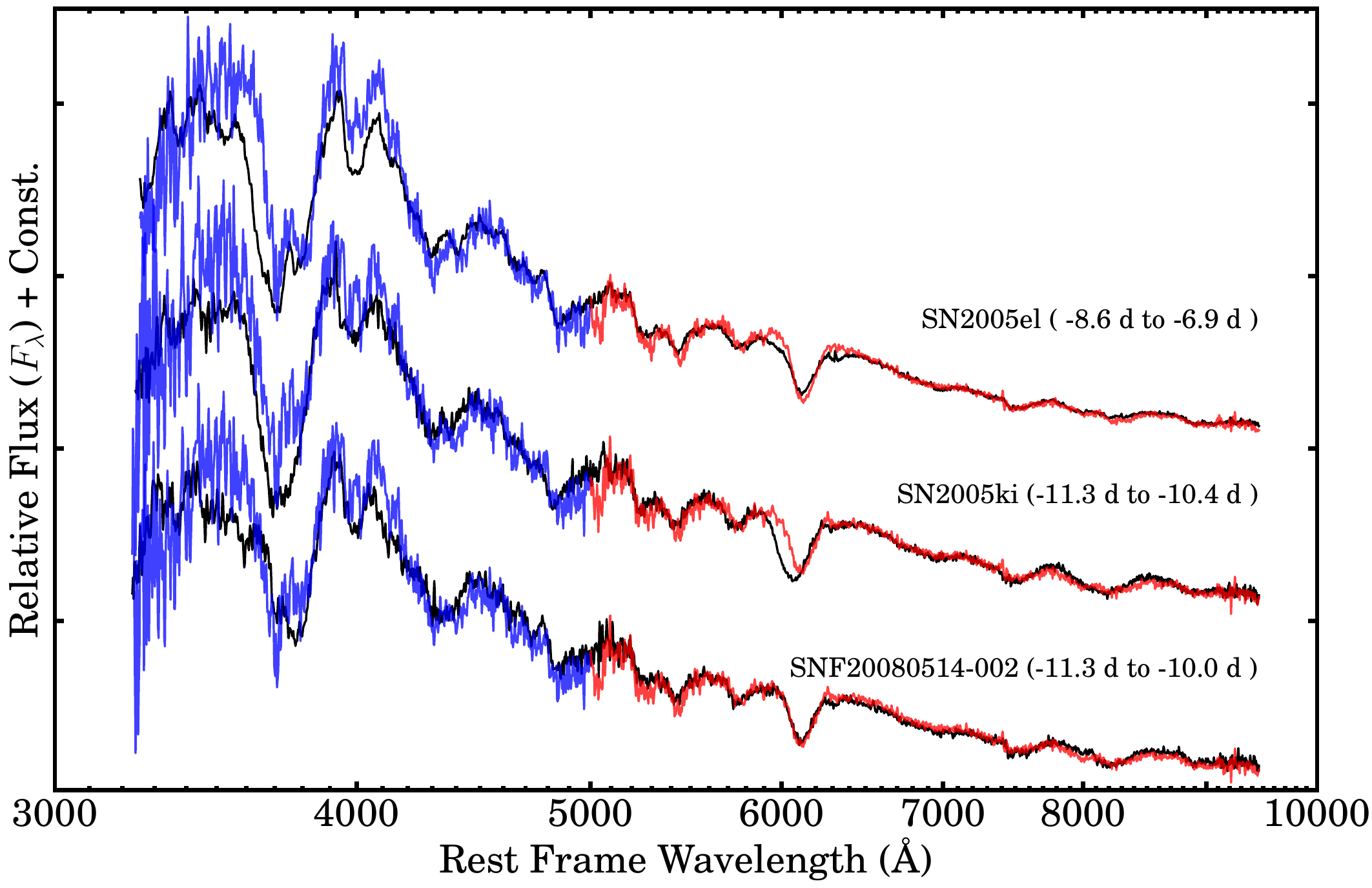}
    \caption
    {
        Selected spectra extracted from time series of SN~2005di, warped
        by a CCM dust law to match the time series of SNe~2005el, 2005ki,
        and SNF20080514-002.  SN~2005di at $-8.6$~d is compared to
        SN~2005el at $-6.9$~d at top; at $-11.3$~d is compared to
        SN~2005ki at $-10.4$~d in the middle; and at $-11.3$~d to
        SNF20080514-002 at $-10.0$~d.  The red line is the wavelength
        portion of the SN~2005di time-series (above 5000~\AA) where the
        data were warped to match.  The blue line is SN~2005di,
        extrapolated with the derived CCM dust law correction.  The
        black line is a spectrum from the target SN warped to in each
        case.
    }
    \label{fig:matches}
\end{figure*}

It is interesting that 3 of the 5 carbon-positive \sneia\ have
relatively narrow light curves, with low SALT2 $x_1$ values around $-2$
($s \sim 0.87$, or $\Delta m_{15}(B) \sim 1.47$, see
Table~\ref{tab:list}).  These same targets also have bluer than normal
colors.  SNF20080514-002 is also among the \sneia\ observed by
Swift/UVOT \citep{gehrels2004, roming2005} with largest UV excesses
\citep{atel1535}, as shown in Figure~\ref{fig:uvot} \citep[adapted
from][]{milne2010}.  Swift/UVOT photometry for SNe~2008hv and
SNF20080514-002 will be presented in Brown \etal\ 2011, in preparation.
Figure~\ref{fig:lc_params} is a plot of SALT2 fit parameters for all
\sneia\ with SNIFS light curve fits and a spectrum before $-5$~d, the
latest phase for which any \ion{C}{2}~\wl 6580 is positively detected in
this sample.  SNe~2005el, 2005ki, and SNF20080514-002 form a relatively
tight ``cluster'' at the fast-declining (low $x_1$) but blue corner
of this plot.  Other published results for SN~2005el have a somewhat
larger value for $x_1$ \citep{hicken2009b, contreras2010}, though all
measurements indicate that $x_1$ is negative and this SN belongs in this
group of relatively narrow light curves with blue colors.  Thus, the
quantitative results for SN~2005el may be further refined, but our
qualitative conclusions should remain unchanged.

Neither SN~2005cf nor SN~2005di fit into the fast-decliner/blue-color
family.  In the case of SN~2005cf, this may not be surprising since the
\ion{Si}{2} and \ion{Ca}{2} features quite robustly trace out HV ejecta
more dramatically than in the other SNe at the same phases, indicating
real physical differences between SN~2005cf and the others.  However, as
described in \S~\ref{sec:data} there is evidence that SN~2005di is
extrinsically reddened.  Since our data do not resolve the Voigt
profiles of the possible multiple velocity components comprising the
\ion{Na}{1}~D absorption line, we are unable to apply a quantitative
correction for extinction using \ion{Na}{1}~D equivalent width
\citep{poznanski2011}.  Instead, we use a different procedure to
investigate whether SN~2005di could simply be a reddened member of the
fast-decliner/blue-color subclass.

To see if this is at least a plausible hypothesis, we warp the spectral
time series of SN~2005di according to the CCM dust law, fitting for the
optimal differential $E(B-V)$ that brings the spectra into detailed
agreement above 5000~\AA\ with each of SNe~2005el, 2005ki and
SNF20080514-002.  We make no attempt to constrain the differential $R_V$
and we assume a standard Milky Way value of 3.1.  If we assume that the
main difference between SN~2005di and the other SNe is the amount of
dust extinction the former suffers, then correcting for it should bring
the spectral time series into alignment.  In the case of perfect
``twin'' SNe, the agreement above 5000~\AA\ should be exquisite on a
feature-by-feature basis, and the extrapolated flux below 5000~\AA\ a
good match.

The warping technique and a fully developed twin SN study are the
subject of a future paper \citep{fakhouri2011}, here we briefly describe
the mechanics for 2 given objects.  Spectral observations of the objects
are paired based on phase, with similar phases being matched to each
other.  A linear interpolation correction to the fluxes accounts for
overall flux difference arising from the small phase differences.  A CCM
dust law with fixed $R_V = 3.1$ is fit to the ensemble of paired spectra,
as well as an individual scale factor for each pair, to account for any
small scatter.  By fitting all of the pairs (extending from early epochs
to late epochs) jointly for the appropriate dust law warping, we
minimize the influence of any particular features on the fit at any
given phase.

Figure~\ref{fig:matches} shows pairs of spectra that result when the
time series of SN~2005di is warped to match the flux from another SN
above 5000~\AA, roughly matching by phase.  The earliest pair of phases
formed in each of the 3 cases is shown.  The agreement above 5000~\AA\
on a feature-by-feature basis is quite good, though there is a
significant difference in the \ion{Si}{2}~\wl 6355 absorption feature
with respect to SN~2005ki --- clearly an ``intrinsic'' difference.
Blueward of the 5000~\AA\ cut, the extrapolated color looks quite
consistent.  Thus, while we cannot prove definitively that SN~2005di is
a dust-estranged member of the fast-decliner/blue-color family, there does
seem to be a resemblance.  Figure~\ref{fig:lc_params} includes the
position of SN~2005di when corrected for the median derived differential
$E(B-V) = 0.40$ is removed.

\section{Discussion}

\label{sec:discussion}

To summarize, our most important findings are:

\begin{itemize}

\item Definitive new detections of \ion{C}{2}~\wl 6580 absorption
notches in the pre-maximum spectra of 5 largely spectroscopically normal
\sneia, at apparent blueshifts around $12,\!000$~\kms, consistent with
the velocity of the ejecta at the photosphere.  

\item The observed notches fade with time without appreciable shift in
wavelength, disappearing before maximum light.  Detailed fitting shows
that \ion{C}{2}~\wl 6580 influences the formation of the spectrum beyond
the immediate vicinity of the absorption notch.

\item Three of the 6 carbon-positive \sneia\ (including SN~2006D) in the
full SNfactory sample group together in light curve shape and color
space at the fast-decliner/blue-color corner.  A further SN may belong
to this family, plausibly separated from it by the effect of extrinsic
dust reddening.

\end{itemize}

We consider 4 main questions in light of our results.  First, factoring
in the limitations of our sample and approach, what is the incidence of
photospheric carbon in pre-maximum spectra of normal \sneia?  Second, to
what extent does carbon-positivity in \sneia\ predict a fast-declining
light curves and blue color (or vice-versa), and what is the physical
significance of such a connection?  Third, what inferences can we make
about the geometry of the ejected carbon in the \sneia\ studied here?
Finally, what impact do our findings have on our understanding of the
\snia\ explosion mechanism?

\subsection{Incidence of \ion{C}{2}}

Our data suggest that unburned carbon, while far from ubiquitous, is
also not particularly rare --- in line with the result of
\citet{parrent2011}.  A detailed rate calculation is beyond the scope of
this article, but we can try to estimate the frequency of
photospheric-velocity carbon in normal \sneia\ based on the latest phase
($-5.0$~d) and lowest median S/N per 2.4~\AA\ bin in the wavelength
range of interest (11.7, that of SN~2005di at $-11.3$~d) where
\ion{C}{2}~\wl 6580 is detected.  31 \sneia\ pass these cuts in the
full SNfactory sample \citep[which includes SN~2006D,][]{thomas2007}, of
which 2 are super-Chandrasekhar candidates and a further 2 are
SN~1991T-like.  Excluding these 4 SNe places the fraction of \sneia\
with \ion{C}{2}~\wl 6580 that persists to $-5$~d at $22^{+10}_{-6}$\%
\citep[Bayesian binomial confidence interval, see][]{cameron2010}.

\subsection{Carbon in \sneia}

\begin{figure*}[htbp]
    \centering
    \includegraphics[width=0.49\textwidth,clip=true]{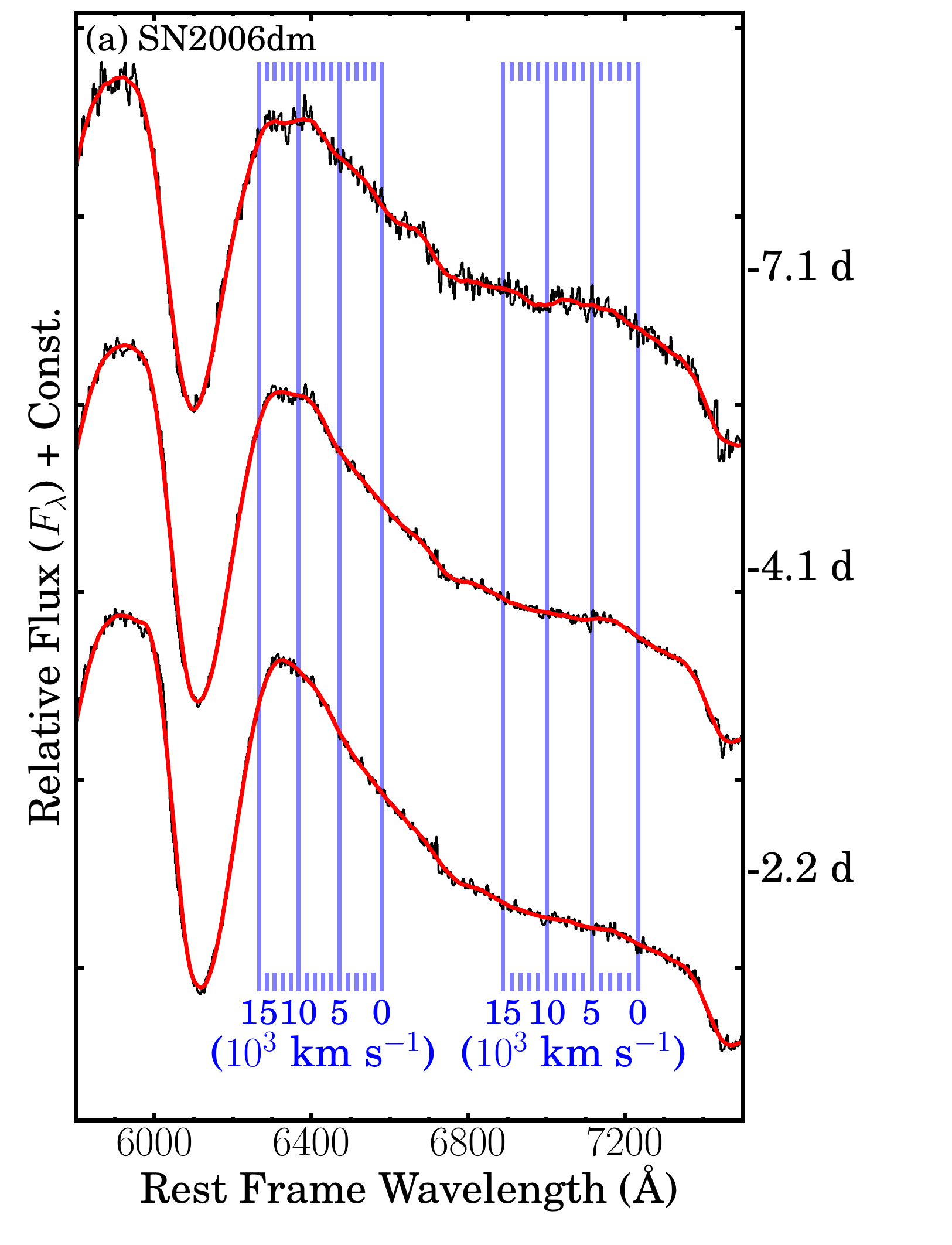}
    \includegraphics[width=0.49\textwidth,clip=true]{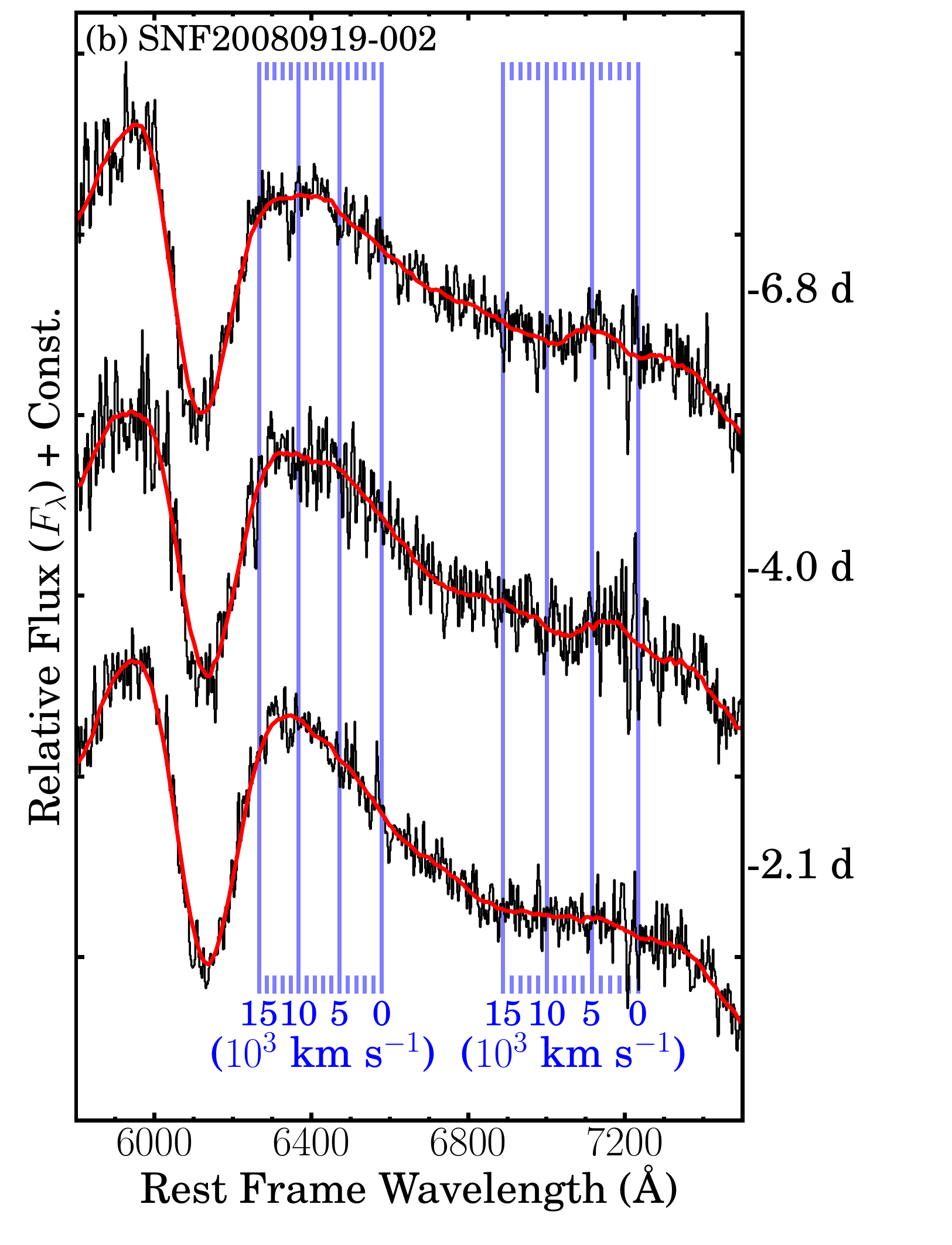}
    \caption
    {
        \ion{C}{2}~\wl 6580 feature regions for SN~2006dm and
        SNF20080919-002, 2 \sneia\ with pre-maximum coverage and light
        curve widths similar to those of SNe~2005el, 2005ki, and
        SNF20080514-002; however these objects are redder.  Distinct
        \ion{C}{2}~\wl 6580 notches are not obviously visible in these
        objects, but the apparent \ion{Si}{2}~\wl 6355 emission peaks
        appear flattened early on.
    }
    \label{fig:other_carbon}
\end{figure*}

SN~2005el, SN~2005ki, SNF20080514-002, and perhaps SN~2005di could be
members of a physically meaningful family of \sneia\ characterized by
detectable carbon in their early spectra, relatively fast-declining
light curves, and intrinsically blue colors.  These 4 SNe were selected
solely based on the presence of carbon signatures in their spectra, so
we now reverse the selection and re-examine the early spectra of other
objects with similar light curve or color parameters.  

The most interesting results from our surveyed sample come from objects
with similar light curve width.  Figure~\ref{fig:other_carbon} shows the
5800--7500~\AA\ region of the pre-maximum spectra of SN~2006dm
\citep{schwehr2006} and SNF20080919-002.  Neither case presents a robust
notch, however the \ion{Si}{2}~\wl 6355 emission peak looks somewhat
``squashed'' in the earliest spectra.  The evidence is hardly decisively
in favor of photospheric-velocity carbon in these 2 additional \sneia,
but it is tantalizing.  Carbon-positive SN~2006D \citep{thomas2007} is
also plotted in Figure~\ref{fig:lc_params}.  It also has a SALT2 $x1$
around $-2$ but is redder than the objects uncovered by our survey here.

Returning to the Swift/UVOT color-curves in Figure~\ref{fig:uvot}, we
identify SN~2008Q \citep{villi2008} and SN~2008hv \citep{pignata2008} as
having UV colors similar to those of SNF20080514-002.  All 3 of these
targets separate from the rest of the Swift sample especially well in $u
- v$ color space.  While the spectroscopic classification announcements
for neither SN~2008Q \citep{stanishev2008} nor SN~2008hv
\citep{marion2008, challis2008, folatelli2008} mention
\ion{C}{2}~\wl 6580, the pre-maximum spectra of both objects include
the absorption notch signature (V. Stanishev and G.~H.~Marion,
private communication).  Both \citet{ganeshalingam2010} and
\citet{brown2010} report a faster-than-average decline rate
(corresponding roughly to $x_1$ of $-0.92$ and $-1.67$ respectively)
for SN~2008Q, suggesting that it could also belong with the
fast-decliner/blue-color family.  Optical light curve analysis for
SN~2008hv is yet to be published, but a preliminary decline-rate
estimate from the UVOT photometry appears to be faster than average.

Next, we consider whether normal \sneia\ with previous \ion{C}{2}~\wl
6580 detections in the literature have a tendency towards narrow light
curves and blue colors.  A revisit of \citet{parrent2011} yields mixed
results, partially because of the lack of consistent light-curve fits
across all the data sets.  Of the 2 ``definite'' cases presented there,
SN~1990N fails to fit the pattern ($x_1 = 0.54 \pm 0.10, c = 0.10 \pm
0.01$); the other is SN~2006D.  The ``probable'' category (where data
are available) is dominated by narrower light curves than average as well,
but no strong tendency toward blue colors.  The exact role that host
galaxy dust extinction plays in the \citet{parrent2011} sample is
unconstrained, however.  Overall, it would seem that the picture for
unburned carbon in \sneia\ may be a complex one, where there are
multiple subclasses and considerable variation between them.

Four of the 5 newly identified carbon-positive \sneia\ may indeed be
representative of a meaningful subclass, but SN~2005cf stands apart.
Its light curve shape is less extreme and its colors more red.  It also
lacks the large UV excess of SNF20080514-002 seen in
Figure~\ref{fig:uvot}.  HV
ejecta signatures in the spectra of SN~2005cf are much stronger than in
the other carbon-positive \sneia\ considered here.  Though HV
\ion{Ca}{2} features may be ``ubiquitous'' \citep{mazzali2005}, it is
clear that there is considerable variation in the velocities and line
strengths observed.  Cases with the strongest HV \ion{Ca}{2} IR triplet
also seem to exhibit strong \ion{Si}{2}~\wl 6355 absorption early on
\citep{stanishev2007}.  The origin of the HV features remains unclear
\citep[Is it a density or composition enhancement in the ejecta, or is
it swept-up circumstellar material?  See e.g.,][]{gerardy2004, tanaka2008,
wang2009b}.  Evidence for a connection between carbon and the HV
features is weak.  The best candidate aside from SN~2005cf with HV
features and a possible \ion{C}{2}~\wl 6580 notch seems to be SN~2003du
\citep{stanishev2007, tanaka2008}.  On the other hand, SN~2009ig (also
with HV features), appears to be a counter-example \citep[not a distinct
notch, so][label it at most a ``possible'' detection]{parrent2011}.
SNF20071021-000, the non-detection example in Figure~\ref{fig:others} is
yet another.

A number of ``peculiar'' \sneia\ (or \snia-like objects) with carbon
lines have been discussed in the literature.  Here we compare our sample
of normal \sneia\ with those objects.

\ion{C}{2} features are a spectroscopic bellwether of candidate
super-Chandrasekhar \sneia, where they seem to persist to maximum light
and even beyond \citep{howell2006, hicken2007, scalzo2010, yamanaka2009,
silverman2010}.  This has been interpreted as evidence supporting the
super-Chandrasekhar hypothesis, where more total carbon is present in
the progenitor and/or much remains after disruption.  These objects are
quite overluminous, and have much broader than normal light curves.
Their spectra, while mostly consistent with one another, are quite
different from those of objects considered here.  Overall, if the
super-Chandrasekhar candidates form a subclass of \sneia, it is
distinct from the one proposed here.

With varying degrees of certainty, carbon is also occasionally detected
in low-luminosity, short time-scale, or otherwise peculiar \sneia.
\citet{taubenberger2008} claim that \ion{C}{2}~\wl 6580 is present in
the spectra of SN~2005bl, a low-luminosity SN~1991bg-like \snia, but
carbon is seldom remarked upon or observed in other members of this
subclass.  More unusual events like SN~2002cx \citep{li2003,
branch2004}, SN~2005hk \citep{chornock2006, phillips2007}, SN~2007qd
\citep{mcclelland2010}, and SN~2008ha \citep{foley2009, foley2010a,
valenti2009} have been conjectured to represent a distinct population of
peculiar, low-luminosity \sneia.  These SNe are quite sub-luminous (peak
$M_B \gtrsim -18$), have light curves that evolve more rapidly than
normal \sneia, and have spectra that exhibit many narrow absorption
lines that suggest lower kinetic energies than typical.  Pure white
dwarf deflagrations or failed deflagrations may provide a plausible way
to account for those observations, in which case carbon signatures may
also be expected \citep[e.g.,][]{ropke2007}.  The carbon identifications
in the spectra of these events are irregular, tentative in most cases,
though SN~2008ha presents the most clear \ion{C}{2} signatures
\citep{foley2010a}.  Spectra of candidate ``SNe~.Ia''
\citep{bildsten2007} SN~2002bj \citep{poznanski2010} and SN~2010X
\citep{kasliwal2010} exhibit robust \ion{C}{2} detections, but so far
detailed simulations of SNe~.Ia \citep{shen2010} do not predict
observable carbon.  All of the above cases are much dimmer than the SNe
considered here, and much more exotic spectroscopically.

SN~2006bt \citep{foley2010b}, however, merits specific attention.  An
outlier on the \snia\ light-curve-width/luminosity relation, SN~2006bt
exhibits some spectroscopic peculiarities.  Most interestingly, its
earliest spectra possess a unique \ion{C}{2}~\wl 6580 signature, a notch
centered at $5,\!200$~\kms, apparent in at least 2 consecutive
observations.  There is little, if any, evidence for emission to the red
of the notch, and the vast difference between the apparent blueshift of
the \ion{C}{2}~\wl 6580 notch and the velocities of other ejected
elements ($12,\!500$~\kms\ for \ion{Si}{2}~\wl 6355) is quite
striking.  \citet{foley2010b} suggest that the discrepancy can be
accounted for by an ejected blob moving at an angle to the line of
sight.  SN~2006bt at the present seems to be the strongest evidence for
carbon ejecta distributed nonspherically in \sneia.

\subsection{Geometry}

\label{sec:geometry}

\begin{figure}[htbp]
    \centering
    \includegraphics[width=0.49\textwidth,clip=true]{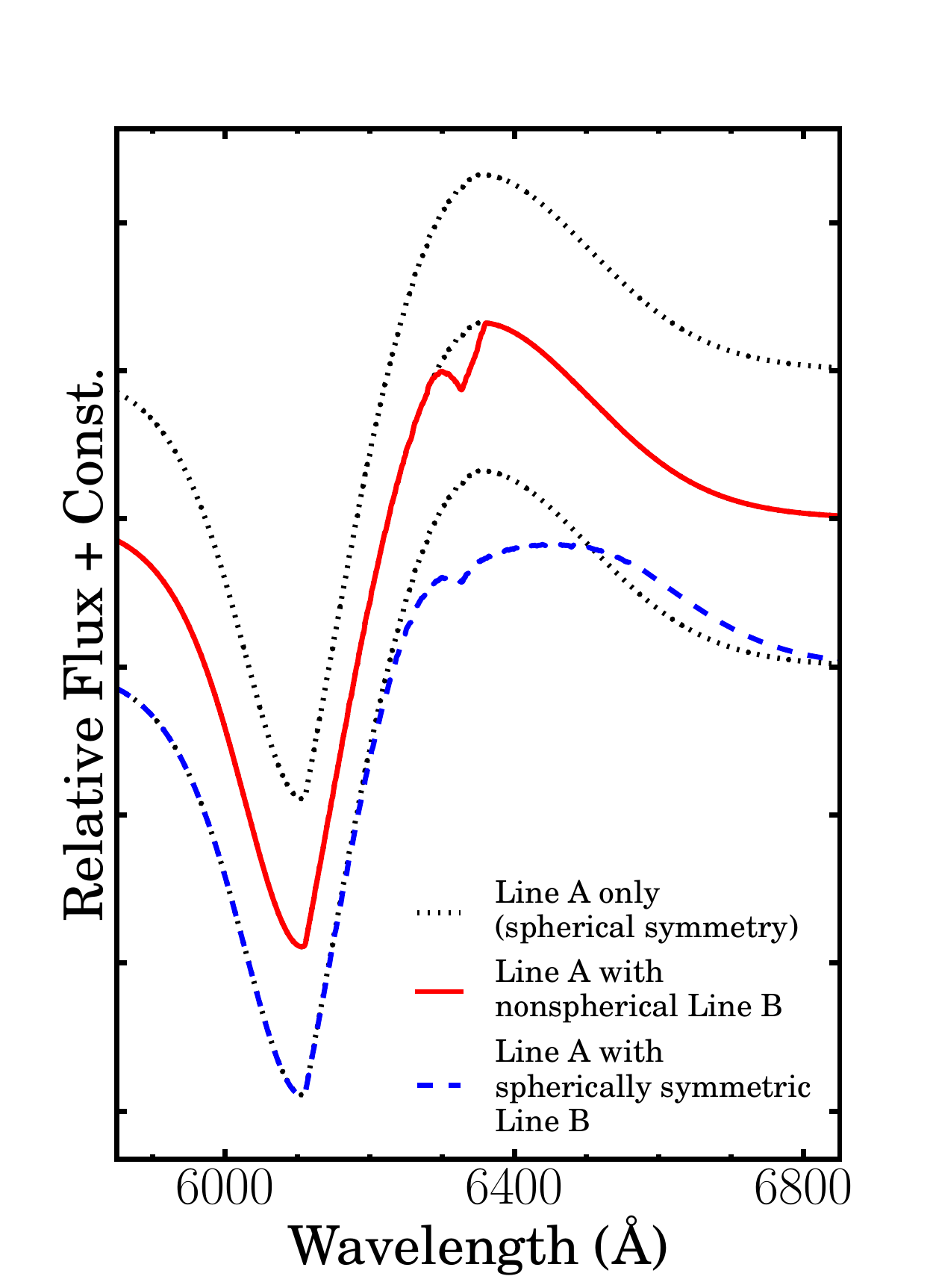}
    \caption
    {
        Simulated, blended line profiles for different distributions of opacity
        and line strength.  The black dotted curve traces a baseline spectrum,
        arising from a single unblended line (``line A'') at rest wavelength
        6355~\AA.  In all cases, line A opacity is spherically symmetric.  The
        red/solid curve is the spectrum resulting when line A is blended with
        another line (``line B'' 6580~\AA) and line B opacity is distributed in
        a blob only partially blocking the photosphere.  The blue/dashed curve
        shows the blend with line B opacity distributed in spherical symmetry,
        and demonstrates the screening effect observed in the carbon-positive
        \sneia\ discussed here.  The red/solid and blue/dashed curves are
        overlaid on the unblended line A profile for visual reference, as was 
        done with the synthetic spectra with and without \ion{C}{2} in
        Figure~\ref{fig:closeup}.
    }
    \label{fig:blending}
\end{figure}

Inspired by the characteristics of multi-dimensional explosion
simulations with strong asymmetries, observers have searched for
spectroscopic and spectropolarimetric indications of deviation from
spherical composition symmetry in \sneia.  Spectropolarimetry offers the
best means for probing the geometry of SN ejecta, but opportunities for
follow-up are relatively rare and the available sample is small.
\citet{wang2008} describe spectropolarimetric observations of 2 \sneia\
in our carbon-positive sample, SNe~2005cf and 2005el (a single epoch each).
In neither case is there a particularly strong case for deviation from
basically spherical composition symmetry.  A dominant axis in the Stokes
Q/U diagram is at least identifiable for SN~2005el, suggestive of some
combination of global and smaller scale asymmetries.

While spectropolarimetry may be more definitive for constraining the
geometry of the unburned material, it is interesting to consider the
constraints we can set with our pure flux measurements.  The comparison
of synthetic spectra with and without \ion{C}{2} shown in
Figure~\ref{fig:closeup} suggests that the spatial distribution of
unburned carbon in the ejecta may be quite consistent with spherical
symmetry.  Figure~\ref{fig:blending} demonstrates our reasoning simply,
using toy model synthetic line profiles.  

The classic SN spectroscopic feature formed by a single line is plotted
at the top of Figure~\ref{fig:blending}: Blueshifted absorption produced
by material in front of the photosphere and expanding toward the
observer, with an emission hump centered at the rest wavelength of the
line arising from material adjacent to the photosphere in projection.
For discussion purposes, we label this line ``line A'' and assign it a
rest wavelength of 6355~\AA.  

Consider a second line, ``line B'' with rest wavelength 6580~\AA, a
Doppler shift of $12,\!000$~\kms, a shift quite similar to the
cases seen in the data.  If the ejecta bearing the parent ions of line B
are distributed in a ``blob'' that only partially covers the photosphere
and/or line A emissivity in projection, it is quite easy to construct a
line B absorption notch blended with the line A emission component.  An
example of the resulting blend appears in the middle of
Figure~\ref{fig:blending}, overlaid on the line A profile.

Now, suppose the line B ions are distributed spherically symmetrically.
In this case, a quite different blended profile results.  Line B notches
can be produced, but not without additional disruption of the line A
profile.  Flux redward of the notch is suppressed and there is even a
mild red emission feature relative to the single unblended line A case.
Such a blend is depicted at the bottom of Figure~\ref{fig:blending},
again overlaid on the single line A profile for comparison.

Because line B opacity does not encircle the photosphere in the blob
case, its effect is restricted to wavelengths corresponding roughly to
where the blob is optically thick in blueshift parallel to the line of
sight.  However, when line B ions are distributed under spherical
symmetry, the photosphere is cocooned by them, and they scatter away
continuum and line A photons alike over a larger range of projected
velocity.  This results in the notch, the flux suppression to the red of
it, and the small emission further on.  Essentially, the driving factor
is the projected area over which line B photons screen continuum but
also line A emissivity, at a given wavelength.

The fits in Figure~\ref{fig:closeup} appear more consistent with the
spherically symmetric case demonstrated in Figure~\ref{fig:blending}.
Our toy models are quite simple, but illustrate the leading-order
screening effect quite clearly.  At the very least, our results
highlight the critical need for detailed spectroscopy and
spectropolarimetry of \sneia\ at early phases to make the most
definitive conclusions about the geometry of the carbon ejecta.

\subsection{Implications for Explosion Models}

The possible existence of a family of \sneia\ with similar properties to
those demonstrated by SN~2005el, SN~2005ki, SNF20080514-002, and perhaps
SN~2005di, is particularly intriguing.  The combination of a narrow 
light curve, large UV excess or at least blue colors, and the presence of
photospheric velocity carbon suggests a very particular conceptual
picture.  A faster than average decline rate indicates lower than average
luminosity for a \snia, in turn suggesting a lower than average mass of
freshly synthesized radioactive nickel.  The blue colors suggest that
the amount of iron-peak elements synthesized or mixed into the outer
layers is low (suppressing line-blanketing at shorter wavelengths), and
these elements are limited to lower velocities.  Also, a relatively high
carbon abundance in the material above the photosphere indicates less
efficient processing of the white dwarf progenitor material.

The above picture is somewhat consistent with a subset or region of
delayed-detonation model space.  Estimates of the mass fraction of
photospheric carbon in \sneia\ are generally low \citep[of order 1\%,
e.g.][]{branch2003, marion2006, thomas2007, tanaka2008, maeda2010a}, and
are also generally consistent with most such models.  Of particular
interest is the recent grid of multi-dimensional models presented by
\citet{kasen2009}.  In these models, a strong ignition (arising from
numerous initiating sparks) raises the amount of burning that takes
place during the initial sub-sonic deflagration phase.  This allows the
outer, unburned layers to pre-expand more relative to the weaker
ignition case.  When the later detonation wave passes through the
pre-expanded material, nuclear burning is less efficient and the
abundance of iron-peak elements is lower relative to the weak ignition
case.  This results in relatively lower radioactive nickel mass which is
more centrally peaked in the ejecta.

However, the total amount of unburned carbon left in these models is
below 1\% by mass, and the S/N of the synthesized Monte Carlo spectra is
too low to attempt to detect any carbon signatures in them.  This result
is actually at variance to earlier (one-dimensional) delayed-detonation
models \citep[e.g.,][]{hoflich2002}, where some amount of unburned
carbon can be left behind.  Simply put, the multi-dimensional model
grids may not yet be extensive enough.

In most of the carbon-positive \sneia\ discussed here, there seems to be
intriguing evidence that the carbon is distributed roughly spherically
symmetrically, and at velocities overlapping silicon.  There may still
be room for deviation from spherical symmetry in the distribution of
unburned material that we cannot rule out here.  For example, large
scale asymmetries, such as proposed by \citet{maeda2010b} to explain
spectroscopic diversity, may allow for almost hemispherical covering of
the photosphere by carbon.  In that particular picture, an offset
ignition in the white dwarf progenitor yields an asymmetric distribution
of burning products.  This scenario is proposed to explain the origin of
``high-velocity gradient'' (HVG) and ``low-velocity gradient'' (LVG)
subclasses of \sneia, segregated by the rate of decrease in measured
\ion{Si}{2}~\wl 6355 blue-shift with time \citep{benetti2005}.  Viewed from the
side closest to the ignition, one observes an LVG \snia; and viewed from the
other an HVG \snia.  Detonation products are distributed in the outer
layers in all directions, but it is not yet clear how complete the
burning is.  Interestingly, \citet{tanaka2008} remarked that of 3 LVG
\sneia\ they studied, all had marginally detected \ion{C}{2}~\wl 6580,
and blue colors.  Their analysis suggests that LVG \sneia\ undergo less
intense burning than their counterparts.  A systematic examination of
\ion{Si}{2}~\wl 6355 blueshift evolution of SNfactory \sneia\ is in
progress, but our preliminary results indicate that all 5 objects
identified here are also LVG members.  Whether or not a
\citet{maeda2010b} scenario is capable of explaining all the
observables, and can yield carbon amounts and distributions consistent
with observations, remains to be seen through detailed multi-dimensional
explosion modeling and radiative transfer.

\section{Conclusion}

\label{sec:conclusion}

We have presented new detections of carbon signatures in the spectra of
several \sneia.  For the most part, the carbon-positive SNe are normal
spectroscopically, with the only significant deviation being the
presence of HV \ion{Ca}{2} and \ion{Si}{2} in one of them.  Four of the
5 objects discussed may form a robust subfamily of \sneia\ with a
relatively fast decline rate and intrinsically blue colors.  The carbon
is detected at velocities consistent with the ejection velocity at the
photosphere, and in the same velocity range as other freshly synthesized
material.  The distribution of carbon is apparently consistent with
spherical or near-spherical symmetry.  The behavior of most of the
carbon-positive \sneia\ discussed here suggests strong radial abundance
stratification, consistent with some delayed detonation models.

Only one of the objects uncovered in the course of our analysis has ever
been previously identified in the literature as carbon-positive, either
at the time of initial spectroscopic classification or in subsequent
analysis.  Our intent in pointing this out is not to criticize; indeed
we have highlighted cases that earlier escaped our notice as well.
Rather, we suggest that the \ion{C}{2}~\wl 6580 notch signature is not
well-known, easy to overlook, or easy to mistake for a reduction
artifact.  In fact, this is part of why we required 2 spectra for labeling
detections as definitive.  The examples provided here should help to sensitize
observers to carbon signatures in early-phase \snia\ spectra.

Understanding the frequency, distribution, and amount of unburned carbon
in \sneia\ is indeed a useful discriminator among competing explosion
models, so constraints on theory depend on reliable and efficient early
triggering of spectroscopic (and spectropolarimetric) follow-up of
\sneia\ for detailed study.  As current \snia\ cosmology efforts
approach their limits in terms of calibration and empirical
standardization, enhancing our understanding of \snia\ physics becomes
all the more important.  Wide-field and high-throughput robotic surveys
have the power to grow the early-time data set, when coupled efficiently
with ample spectroscopic follow-up resources.  This opens new
opportunities for fully exploiting the analysis of the outer layers of
\snia\ ejecta to constrain explosion models.

\acknowledgements

We thank the anonymous referee for carefully reading our manuscript and
for providing a thorough report that improved our paper's clarity.  We
are grateful for assistance from the technical/scientific staffs of
the Palomar Observatory, the High Performance Wireless Radio Network
(HPWREN), the National Energy Research Scientific Computing Center
(NERSC), and the University of Hawaii 2.2~m telescope.  We also wish
to recognize and acknowledge the significant cultural role and
reverence that the summit of Mauna Kea has always had within the
indigenous Hawaiian community.  We are most fortunate to have the
opportunity to conduct observations from this mountain.  This work
was supported by the Director, Office of Science, Office of High
Energy Physics, of the U.S.  Department of Energy under Contract
No.~DE-AC02-05CH11231; the U.S.  Department of Energy Scientific
Discovery through Advanced Computing (SciDAC) program under Contract
No.~DE-FG02-06ER06-04; by a grant from the Gordon \& Betty Moore
Foundation; in France by support from CNRS/IN2P3, CNRS/INSU, and
PNC; and in Germany by the DFG through TRR33 ``The Dark Universe.''
Funding was also provided by a Henri Chretien International Research
Grant administrated by the American Astronomical Society; the
France-Berkeley Fund; and by an Explora'Doc Grant by the Region
Rhone Alpes.  Much of this research was conducted with resources and
support from NERSC, which is supported by the Director, Office of
Science, Office of Advanced Scientific Computing Research, of the
U.S.  Department of Energy under Contract No.~DE-AC02-05CH11231.
HPWREN is funded by National Science Foundation Grant Number
ANI-0087344, and the University of California, San Diego.  We
appreciated constructive discussions with D.~Kasen, J.~Parrent,
D.~Poznanski and R.~Foley.  We thank V.  Stanishev for sharing a new
reduction of a spectrum of SN~2008Q with us in advance of
publication.  We thank G.~H.  Marion, R.~P.  Kirshner, and P.
Challis (Harvard-Smithsonian Center for Astrophysics) for sharing
unpublished spectra of SN~2008hv.  The CfA Supernova Program is
supported by NSF grant AST~09-07903.  P.~A.~M. acknowledges support
from the NASA Astrophysics Data and Analysis Program Grant
NNX06AH85G.  R.~C.~T.  thanks fellow participants at the Aspen Center
for Physics Summer 2010 workshop ``Taking Supernova Cosmology into the
Next Decade,'' for their input.

\appendix

\section{Notes on Individual Supernovae}

\subsection{SN~2005di}

\textbf{\ion{C}{2}.}  The \ion{C}{2}~\wl 6580
absorption notch is clearly visible in the $-11.3$~d spectrum, located
between roughly $9,\!000$ and $15,\!000$~\kms\ with a minimum around
$12,\!000$~\kms\ in apparent blueshift.  Almost 3 days later, the notch
is barely discernible, and it is clearly absent by $-3.6$~d.  In the
last spectrum, it is unlikely that the noise is obscuring a notch: The
change in the shape of the spectrum between the last 2 spectra is overt.
Examining the \ion{C}{2}~\wl 7234 region, we detect an inflection in the
shape of all 3 spectra, but its position is inconsistent across phase.
This inconsistency may be a consequence of the low S/N of the
observations.  In the first spectrum, the flux depression is coincident
with the \wl 6580 notch in blueshift, but this is not the case in the
later spectra.  However, in all 3 spectra, a bump in the flux centered
at a \wl 7234 blueshift of $5,\!000$~\kms\ is clearly visible.
Figure~\ref{fig:carbon}a also clearly shows the narrow rest-frame
\ion{Na}{1}~D absorption, in the shaded rectangle.

\textbf{SYNAPPS Fits.} \ion{C}{2} contributes
to the spectrum on $-11.3$~d and also in the next spectrum on $-8.6$~d,
but on $-3.6$~d \synapps\ finds no need for it.  The \synapps\ velocity
at the photosphere ($v_{phot}$) fits are $12,\!600$, $12,\!200$, and
$11,\!900$~\kms\ respectively.  HV \ion{Ca}{2} opacity is invoked (at $v
> 20,\!000$~\kms) though the signature is rather small --- it is
more important for fitting the IR triplet than the UV H\&K feature.
Ultimately, the case for HV \ion{Si}{2} is uncompelling, which is not
surprising since the blue side of the \ion{Si}{2}~\wl 6355 absorption
only extends to $18,\!000$~\kms, in contrast to instances where HV
\ion{Si}{2} has been previously discussed \citep[e.g.,
SN~2005cf,][]{garavini2007} and found to extend upwards of 
$20,\!000$~\kms.

\subsection{SN~2005el}

\textbf{\ion{C}{2}.} A \ion{C}{2}~\wl 6580
absorption notch is readily seen in the $-6.9$~d spectrum.  As in
SN~2005di, the notch is located in the apparent blueshift interval
$10,\!000$ to $15,\!000$~\kms.  There is a clear depression in the flux
around 6980~\AA, possibly attributable to \ion{C}{2}~\wl 7234
absorption, though the blueshift is not completely consistent with that
for the \wl 6580 notch.  Both features disappear from the spectrum by
the time of maximum light.  As in SN~2005di, the red side of the
\ion{Si}{2} emission feature steepens as the SN ages and the
\ion{C}{2}~\wl 6580 dissipates.  There is also an emission bump to
the red of the \wl 7234 notch, centered at a rest-frame blueshift of
$5,\!000$~\kms.  Like the 2 notches, this bump disappears by
maximum light.

\textbf{SYNAPPS Fits.} \synapps\ requires
\ion{C}{2} only in the $-6.9$~d spectrum, and not at maximum light.  The
value of $v_{phot}$ changes from $12,\!500$ to $12,\!000$~\kms\ from
the first spectrum to the second.  HV \ion{Ca}{2} and \ion{Si}{2}
opacity components were enabled as was done with SN~2005di.  The results
were similar, with some HV \ion{Ca}{2} contributing to the IR triplet
and the bluer feature of the H\&K absorption blend, and HV \ion{Si}{2}
making only a minor contribution to the blue wing of the \ion{Si}{2} \wl
6355 absorption in the first spectrum.

\subsection{SN~2005ki}

\textbf{\ion{C}{2}.} In the $-10.4$~d
spectrum, the \ion{C}{2}~\wl 6580 notch appears slightly more
blueshifted than in the first spectra of SNe~2005di and 2005el,
extending just beyond $15,\!000$~\kms.  The same is true for
\ion{Si}{2}~\wl 6355 absorptions.  As
before, the carbon signatures dissipate as the SN brightens: Weaker by
$-8.5$~d and absent by $-5.5$~d.  Also, in the first spectrum, an
emission plateau just redward of the notch is visible, asymmetric with
respect to the rest wavelength of \wl 6580, and again it weakens in
parallel with the notch.  A clear association for \ion{C}{2}~\wl 7234 in
this spectrum is quite problematic: The fading bump at 7100~\AA\ is
again visible, but there is no clear absorption notch.

\textbf{SYNAPPS Fits.} \ion{C}{2} opacity is
used in the $-10.4$ and $-8.5$~d spectra, but not in the $-5.5$~d
spectrum.  \synapps\ photospheric velocities for SN~2005ki are a bit
higher than in the other SNe considered here, with $v_{phot}$ fit at
$13,\!100$, $12,\!600$, and $12,\!500$~\kms.  One major difference is that a
distinct HV \ion{Ca}{2} opacity component is \emph{not} required to
reproduce the \ion{Ca}{2} features, especially at $-10.4$~d: A single
opacity profile extending to high velocity is sufficient.

\subsection{SNF20080514-002}

\textbf{\ion{C}{2}.} The \ion{C}{2}~\wl
6580 notch (apparent blueshift interval again $10,\!000$ to
$15,\!000$~\kms) is most apparent at $-10.0$ and $-7.9$~d, still
discernable at $-5.0$~d, and probably absent at $-3.1$~d.  The notch is
accompanied by a modest and fading emission plateau to the red which is
almost symmetric about the line rest wavelength.  The \wl 7234 blueshift
region is again characterized by an inflection and bump centered
blueward of 7234~\AA, and again the strength of this bump decreases with
time.

\textbf{SYNAPPS Fits.} \ion{C}{2}
opacity is needed in the $-10.0$, $-7.9$, and $-5.0$~d fits, but not on
day $-3.1$.  The values of $v_{phot}$ fit are $12,700$, $12,400$,
$12,100$ and $12,100$ ~\kms, respectively.  Only a modest amount of HV
\ion{Ca}{2} opacity is used in these fits, and HV \ion{Si}{2} proves to
be unnecessary.  

\subsection{SN~2005cf}

\textbf{\ion{C}{2}.}  The \ion{C}{2}~\wl 6580
absorption notch shape is significantly different from that seen in
the other carbon-positive \sneia\ studied here.  The notch is more
narrow, extending in apparent blueshift from $10,\!000$ to
$13,\!000$~\kms\ in the $-9.4$~d spectrum.  The notch as faded away by
$-7.4$~d, a phase at which it is still present in the other
carbon-positive SNe studied here.

\textbf{SYNAPPS Fits.} \synapps\ uses
\ion{C}{2} for the $-9.4$~d spectrum but not for the spectrum obtained a
week later.  \synapps\ determines $v_{phot} = 12,\!800$ and then
$12,\!000$~\kms, respectively.  The $-9.4$~d spectrum requires both HV
\ion{Si}{2} and \ion{Ca}{2}, though the HV \ion{Si}{2} feature becomes
optically thin by $-2.4$~d.  The HV component is also clearly quite
important for the formation of the \ion{Ca}{2} H\&K absorption in both
spectra.  The strength of the HV features in SN~2005cf distinguishes it
from the other \sneia\ considered here.

\end{document}